\documentclass[aps,twocolumn,showpacs]{revtex4}
\usepackage{color}
\usepackage{graphicx}
\usepackage{graphics}
\usepackage{epsfig}
\usepackage{subfigure}
\usepackage{latexsym}
\usepackage{amsfonts}
\usepackage[english]{babel}
\usepackage[latin1]{inputenc}
\usepackage[all]{xy}
\usepackage{amsmath}
\usepackage{amssymb}
\newcommand{\be}{\begin{equation}}
\newcommand{\ee}{\end{equation}}
\newcommand{\ben}{\begin{eqnarray}}
\newcommand{\een}{\end{eqnarray}}
\newcommand{\bes}{\begin{subequations}}
\newcommand{\ees}{\end{subequations}}

\newcommand{\sech}{{\rm sech}}

\def \sech{\mathop{\rm sech}\nolimits}

\def \sech{\mathop{\rm sech}\nolimits}
\def \su{\mathop{\rm su}\nolimits}
\def \sv{\mathop{\rm sv}\nolimits}
\def \sf{\mathop{\rm sf}\nolimits}

\def \cu{\mathop{\rm cu}\nolimits}
\def \cv{\mathop{\rm cv}\nolimits}
\def \cf{\mathop{\rm cf}\nolimits}

\def \tfmedios{\mathop{\rm t\frac{f}{2}}\nolimits}

\newcommand{\ba}[1]{\begin{array}{#1}}
\newcommand{\ea}{\end{array}}
\newcommand{\bea}[1]{\begin{equation}\left\{\begin{array}{#1}}
\newcommand{\eea}{\end{array}\right.\end{equation}}

\begin{document}

\title{\bf BPS and non-BPS kinks in a massive non-linear ${\mathbb S}^2$-sigma model}

\author{A. Alonso-Izquierdo$^{a}$, M. A. Gonzalez Leon$^{a}$ and J. Mateos Guilarte$^{b}$}
\affiliation{{$^{a}$ Departamento de Matematica Aplicada and IUFFyM,
Universidad de Salamanca, SPAIN}
\\{$^{b}$Departamento de Fisica and IUFFyM, Universidad de Salamanca, SPAIN}}

\begin{abstract}
The stability of the topological kinks of the non-linear ${\mathbb
S}^2$-sigma model discovered in \cite{AMAJ} is discussed by means of
a direct estimation of the spectra of the second-order fluctuation
operators around topological kinks. The one-loop mass shifts caused
by quantum fluctuations around the these kinks are computed using
the Cahill-Comtet-Glauber formula \cite{CCG}. The (lack of)
stability of the non-topological kinks is unveiled by application of
the Morse index theorem. These kinks are identified as non-BPS
states. There are two types of topological kinks coming from the
twofold embedding of the sine-Gordon model in the massive non-linear
sigma model. It is shown that sine-Gordon kinks of only one type
satisfy first-order equations and are accordingly BPS classical
solutions. Finally, the interplay between instability and
supersymmetry is explored.
\end{abstract}
\maketitle
\section{Introduction}

The main theme in this paper is the analysis of the structure of the
manifold of kink solitary waves discovered in \cite{AMAJ}. In
particular, we shall offer a full description of the stability of
the different type of kinks. As a bonus, we shall gain information
about the semi-classical behavior of such kinks from the stability
analysis, providing us with enough data to compute the one-loop mass
shifts for the topological kinks.

Prior to our work \cite{AMAJ}, kinks in massive non-linear sigma
models have been known for some time and profusely studied in
different supersymmetric models under the circumstance that all the
masses of the pseudo Nambu-Goldstone particles are equal. The study
started with two papers by Abraham and Townsend \cite{AT},
\cite{AT1} in which the authors discovered a family of Q-kinks in a
(1+1)-dimensional ${\cal N}=(4,4)$ supersymmetric non-linear sigma
model with a hyper-Kahler Gibbons-Hawking instanton as the target
space and mass terms obtained from dimensional reduction. In
\cite{Nit}, however, these kinks were re-considered by constructing
the dimensionally reduced supersymmetric model by means of the
mathematically elegant technique of hyper-Kahler quotients. By doing
this, the authors deal with massive ${\mathbb CP}^N$  or ${\mathbb
HP}^N$ models, a playground closer to our simpler massive ${\mathbb
S}^2$-sigma model. Similar ${\cal N}=2$ BPS walls in the ${\mathbb
CP}^1$-model with twisted mass were described in \cite{Dor}. In a
parallel development in the (2+1)-dimensional version of these
models, two-dimensional Q-lumps were discovered in \cite{Lee} and
\cite{Abr}. Throughout this field, the most interesting result is
the demonstration in \cite{GPTT} and \cite{INOS} that composite
solitons in ${\it d}=3+1$ of Q-strings and domain walls are exact
BPS solutions that preserve $\frac{1}{4}$ of the supersymmetries: (
See also the review \cite{EINOS}, where a summary of these
supersymmetric topological solitons is offered.)

Our investigation differs from previous work in the area of
topological defects in non-linear sigma models in two important
aspects: 1) We remain in a purely bosonic framework; in fact, we
consider the simplest massive non-linear sigma model. 2) We study
the case when the masses of the pseudo Nambu-Goldstone bosons are
different. The search for kinks in the (1+1)-dimensional model
(domain walls in ${\it d}=3+1$) is tantamount to the search for
finite action trajectories in the repulsive Neumann system, a
particle moving in an ${\mathbb S}^2$-sphere under the action of
non-isotropic repulsive elastic forces. It is well known that this
dynamical system is completely integrable \cite{Dub}, \cite{Per}. We
show, however, that the problem is Hamilton-Jacobi separable by
using elliptic coordinates in the sphere. Use of this allows us to
find four families of homoclinic trajectories starting and ending at
one of the poles which are unstable points of the mechanical system.
In the field-theoretical model the poles become ground states,
whereas the homoclinic trajectories correspond to four families of
non-topological kinks. Each member in a family is formed by a
non-linear combination of two basic topological kinks (of different
type) with their centers located at any relative distance with
respect each other.

It is remarkable that the static field equations of this massive
non-linear sigma model are (almost) the static Landau-Lifshitz
equations governing the high spin and long wavelength limit of $1D$
ferromagnetic materials. From this perspective, topological kinks
can be interpreted respectively as Bloch and Ising walls that form
interfaces between ferromagnetic domains, similar to those
discovered in the XY model dealt with in \cite{BW}. The variety of
our non-topological kinks, understood as solitary spin waves, is
thus formed by non-linear superpositions of one basic Bloch wall and
one basic Ising wall at different distances. Far from this
non-relativistic context, degenerate Bloch/Ising branes have been
studied in two-scalar field theories coupled to gravity in \cite{ES,
BG, SAH}.



\section{The (1+1)-dimensional massive non-linear ${\mathbb S}^2$-sigma model}

We shall focus on the non-linear ${\mathbb S}^2$-sigma model studied
in Reference \cite{AMAJ}. The action governing the dynamics is:
\begin{equation}
S[\phi_1,\phi_2,\phi_3]=\int \, dtdx \, \left\{{1\over
2}g^{\mu\nu}\sum_{a=1}^3\frac{\partial\phi_a}{\partial
x^\mu}\frac{\partial\phi_a}{\partial x^\nu} -V\right\} ,
\label{eq:act}
\end{equation}
with $V=V(\phi_1(t,x),\phi_2(t,x),\phi_3(t,x))$. The scalar fields
are constrained to satisfy: $ \phi_1^2+\phi_2^2+\phi_3^2=R^2$, and
thus $\phi_a(t,x)\in {\rm Maps}({\mathbb R}^{1,1},{\mathbb S}^2)$
are maps from the $(1+1)$-dimensional Minkowski space-time to a
${\mathbb S}^2$-sphere of radius $R$, which is the target manifold
of the model.

Our conventions for ${\mathbb R}^{1,1}$ are as follows: $ x^\mu\in
{\mathbb R}^{1,1}$, $\mu=0,1$, $x^\mu\cdot x_\mu=g^{\mu\nu}x_\mu
x_\nu$, $g^{\mu\nu}={\rm diag}(1,-1)$. $x^0=t$, $x^1=x$, $x^\mu\cdot
x_\mu=t^2-x^2$;
$\partial_\mu\partial^\mu=g^{\mu\nu}\partial_{\mu\nu}^2=\Box=\partial_t^2-\partial_x^2$.

The infrared asymptotics of $(1+1)$-dimensional scalar field
theories forbids massless particles, see \cite{Col}. We thus choose
the simplest potential energy density that would be generated by
quantum fluctuations giving mass to the fundamental quanta:
\begin{equation}
V(\phi_1,\phi_2,\phi_3)=\frac{1}{2} \left( \alpha_1^2 \,
\phi_1^2+\alpha_2^2 \, \phi_2^2+\alpha_3^2 \, \phi_3^2\right),
\end{equation}
which we set with no loss of generality such that:
$\alpha_1^2\geq\alpha_2^2
> \alpha_3^2\geq 0$.

\noindent 1. Solving $\phi_3$ in favor of $\phi_1$ and $\phi_2$,
$\phi_3= {\rm sg}(\phi_3) \sqrt{R^2-\phi_1^2-\phi_2^2}$, we find:
\begin{eqnarray}
&&S={1\over 2}\int \, dt dx \,
\left\{\partial_\mu\phi_1\partial^\mu\phi_1+\partial_\mu\phi_2\partial^\mu\phi_2+
\right. \nonumber
\\ && \left.\frac{(\phi_1\partial_\mu\phi_1+\phi_2\partial_\mu\phi_2
)(\phi_1\partial^\mu\phi_1+\phi_2\partial^\mu\phi_2
)}{R^2-\phi_1^2-\phi_2^2}-2V_{{\mathbb
S}^2}(\phi_1,\phi_2)\right\}\nonumber \\
&&V_{{\mathbb S}^2}(\phi_1,\phi_2)=\frac{1}{2} \left(
(\alpha_1^2-\alpha_3^2) \, \phi_1^2+(\alpha_2^2-\alpha_3^2) \,
\phi_2^2+{\rm const.} \right)\nonumber\\ &&\hspace{1cm} \simeq
\frac{\lambda^2}{2}\phi_1^2(t,x)+\frac{\gamma^2}{2}\phi_2^2(t,x)\label{potS2}
\end{eqnarray}
with $\lambda^2=(\alpha_1^2-\alpha_3^2)$,
$\gamma^2=(\alpha_2^2-\alpha_3^2)$, $\lambda^2\geq \gamma^2$.

\noindent 2. Thus, the interactions come from the geometry:
\begin{eqnarray*}
&&\frac{(\phi_1\partial_\mu\phi_1+\phi_2\partial_\mu\phi_2
)(\phi_1\partial^\mu\phi_1+\phi_2\partial^\mu\phi_2
)}{R^2-\phi_1^2-\phi_2^2}\simeq \\
&\simeq& {1\over R^2}\left(1+{1\over R^2}(\phi_1^2+\phi_2^2)+{1\over
R^4}(\phi_1^2+\phi_2)^2+\cdots\right)\cdot\\ &&\cdot\left(
\phi_1\partial_\mu\phi_1+\phi_2\partial_\mu\phi_2\right)\left(
\phi_1\partial^\mu\phi_1+\phi_2\partial^\mu\phi_2\right) \quad ,
\end{eqnarray*}
and ${1\over R^2}$ is a non-dimensional coupling constant, whereas
the masses of the pseudo-Nambu-Goldstone bosons are respectively
$\lambda$ and $\gamma$.

Taking into account that in the natural system of units $\hbar=c=1$
the dimensions of fields, masses and coupling constants are $
[\phi_a]=1=[R]$, $[\gamma]=M=[\lambda]$, we define the
non-dimensional space-time coordinates and masses
\[
x^\mu \, \rightarrow \, \frac{x^\mu}{\lambda} ,\,
\sigma^2=\frac{\alpha_2^2-\alpha_3^2}{\alpha_1^2-\alpha_3^2}=\frac{\gamma^2}{\lambda^2}
,\quad 0 < \sigma^2 \leq 1 ,
\]
to write the energy in terms of them:
\begin{eqnarray}
E=&&{\lambda\over 2}\int \, dx \,
\left\{\left(\partial_t\phi_1\right)^2+\left(\partial_t\phi_2\right)^2+
\left(\partial_x\phi_1\right)^2+ \left(\partial_x\phi_2\right)^2
\right.\nonumber\\ &&
+\frac{(\phi_1\partial_t\phi_1+\phi_2\partial_t\phi_2)^2+(\phi_1\partial_x\phi_1+
\phi_2\partial_x\phi_2)^2}{R^2-\phi_1^2-\phi_2^2}  \nonumber\\
&&+ \left. \phi_1^2(t,x)+\sigma^2\cdot\phi_2^2(t,x)\right\}\
\label{sen}\ \, .
\end{eqnarray}
In the time-independent homogeneous minima of the action or vacua of
our model, $ \phi_1^{V^\pm}=\phi_2^{V^\pm}=0$, $\phi_3^{V^\pm}= \pm
R$ (North and South Poles), the ${\mathbb Z}_2\times{\mathbb
Z}_2\times{\mathbb Z}_2 $, $\phi_a \rightarrow
(-1)^{\delta_{ab}}\phi_b$, $b=1,2,3$ symmetry of the action
(\ref{eq:act}) is spontaneously broken to: ${\mathbb
Z}_2\times{\mathbb Z}_2 $, $\phi_\alpha \rightarrow
(-1)^{\delta_{\alpha\beta}}\phi_\beta $, $ \alpha , \,\beta=1,2$.
Finite energy configurations require:
\begin{equation}
\lim_{x\to\pm\infty}\frac{d\phi_\alpha}{dx}=0 \qquad , \qquad
\lim_{x\to\pm\infty}\phi_\alpha=0 \qquad . \label{febc}
\end{equation}
Therefore, the configuration space ${\cal C}=\left\{{\rm
Maps}({\mathbb R},{\mathbb S}^2)/\right. $ $ \left.E<+\infty
\right\}$ is the union of four disconnected sectors ${\cal C}={\cal
C}_{\rm NN}\bigcup {\cal C}_{\rm SS} \bigcup {\cal C}_{\rm NS}
\bigcup {\cal C}_{\rm SN}$ labeled by the vacua reached by each
configuration at the two disconnected components of the boundary of
the real line.


We now solve the constraint by using spherical coordinates:
$\theta\in [0,\pi]$, $\varphi\in [0,2\pi)$
\begin{eqnarray*}
\phi_1(t,x)&=&R\sin\theta(t,x)\cos\varphi(t,x)\\
\phi_2(t,x)&=&R\sin\theta(t,x)\sin\varphi(t,x)\\
\phi_3(t,x)&=&R\cos\theta(t,x)\, \, \, .
\end{eqnarray*}
In spherical coordinates the mass terms (we shall denote in the
sequel: $\bar{\sigma}=\sqrt{1-\sigma^2}$) are
\begin{equation}
V(\theta,\varphi) = \frac{R^2}{2} \sin^2 \theta (\sigma^2 +
\bar{\sigma}^2 \cos^2\varphi) \qquad ,
\end{equation}
the action becomes
\begin{eqnarray*}
S=\int dtdx &&
\left\{\frac{R^2}{2}\left[\partial_\mu\theta\partial^\mu
\theta+\sin^2\theta \partial_\mu\varphi\partial^\mu
\varphi\right]\right.
\\ &&\left. -\frac{R^2}{2} \sin^2 \theta (\sigma^2 + \bar{\sigma}^2
\cos^2\varphi)\right\} \qquad ,
\end{eqnarray*}
and the field equations read:
\begin{eqnarray}
\Box \theta-{1\over 2}{\rm sin}2\theta\left(\partial^\mu\varphi
\partial_\mu\varphi-\cos^2\varphi
-\sigma^2\sin^2\varphi\right)&=& 0 \label{pfe1}\\
\partial^\mu (\sin^2\theta \partial_\mu\varphi)-{1\over
2}\bar{\sigma}^2\sin^2\theta \sin 2\varphi&=& 0 \, . \label{pfe2}
\end{eqnarray}

Finite energy solutions for which the space-time dependence is of
the form:
\[
\theta(t,x)=\theta\left(\frac{x-vt}{\sqrt{1-v^2}}\right), \quad
\varphi(t,x)=\varphi\left(\frac{x-vt}{\sqrt{1-v^2}}\right) ,
\]
for some velocity $v$, are called solitary waves. Lorentz invariance
allows us to obtain all the solitary waves in our model from
solutions of the static field equations
\begin{eqnarray}
\theta''-{1\over 2}\sin 2\theta \, (\varphi')^2 &=& {1\over 2}
\left(\cos^2\varphi+\sigma^2\sin^2\varphi\right)\sin 2\theta\label{eq:sfe1} \\
\frac{d}{d x}(\sin^2\theta\, \varphi')&=& {1\over 2}\bar{\sigma}^2
\sin^2\theta \sin 2\varphi \label{eq:sfe2} \qquad ,
\end{eqnarray}
where the notation is: $\theta'=\frac{d\theta}{dx}$,
$\varphi'=\frac{d\varphi}{dx}$. The energy of the static
configurations is:
\begin{eqnarray*}
&&E[\theta,\phi]=\lambda \int \, dx \   {\cal E}(\theta^\prime(x),
\varphi^\prime(x), \theta(x),\varphi(x))\   ,\\ && {\cal
E}=\frac{\lambda R^2}{2} \left( (\theta')^2+\sin^2\theta
(\varphi')^2+\sin^2 \theta (\sigma^2 +
\bar{\sigma}^2\cos^2\varphi)\right) \, \, .
\end{eqnarray*}

\section{Topological kinks}

Equation (\ref{pfe2}) is satisfied for constant values of $\varphi$
if and only if: $\varphi=0, \frac{\pi}{2}, \pi, \frac{3\pi}{2}$.
Depending on which pair of $\varphi$-constant solution we choose,
(\ref{pfe1}) becomes one or another sine-Gordon equation:
\[
\Box \theta+{\sigma^2\over 2}\sin 2\theta=0\  ;\quad \Box
\theta+{1\over 2}\sin 2\theta=0 \qquad .
\]
Thus, sine-Gordon models are embedded in our system on these two
orthogonal meridians.

\begin{figure}[htbp]
\centerline{\includegraphics[height=2.8cm]{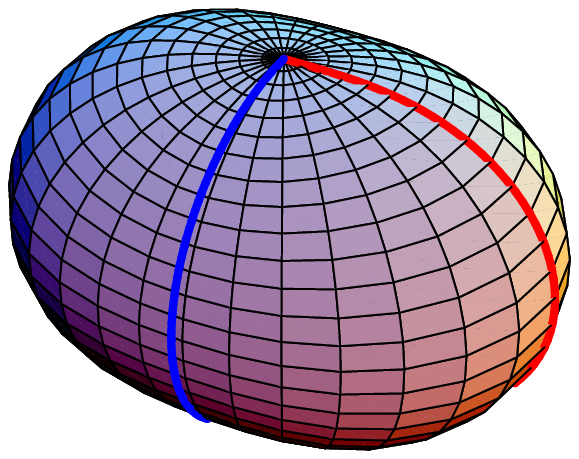} \
\includegraphics[height=3.2cm]{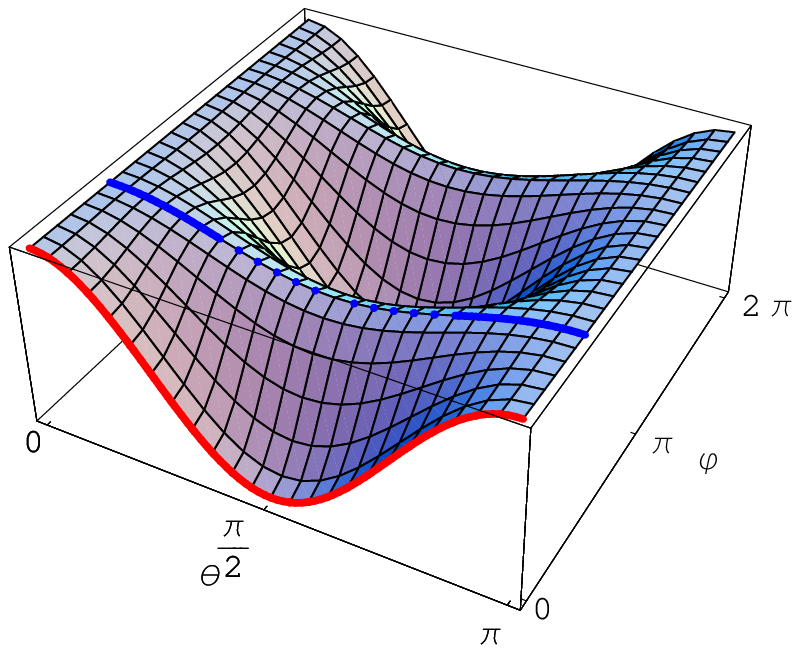}} \caption{a)
 $V(\theta,\varphi)$ deformation of ${\mathbb S}^2$,
b) Embedding of the sine-Gordon model at $\varphi=0$,
$\varphi=\frac{\pi}{2}$ as seen in $-V(\theta,\varphi)$.}
\end{figure}


\noindent {\bf 1. $K_1/K_1^*$ kinks.} We denote $K_1/K_1^*$ the
kink/antikink solutions of the sG model embedded inside the
${\mathbb S}^2$ model in the $\varphi_{K_1}(x)={\pi\over 2}$ or
$\varphi_{K_1^*}(x)={3\pi\over 2}$ two halves of the single meridian
intersecting the $\phi_2:\phi_3$ plane,
\begin{equation}
\theta_{K_1}(x)=\theta_{K_1^*}(x)=2\arctan \, e^{\pm \sigma(x-x_0)}
\qquad , \label{k1kinks}
\end{equation}
see Figure 1. The energy of these kinks, which
belong to ${\cal C}_{\rm NS}$ (kinks) or ${\cal C}_{\rm SN}$
(antikinks), is: $ E^C_{K_1}=E^C_{K_1^*}=2\lambda R^2 \sigma$.


\noindent {\bf 2. $K_2/K_2^*$ kinks.} Taking $\varphi_{K_2}(x)=0$ or
$\varphi_{K_2^*}(x)=\pi$, we find the sG kinks:
\begin{equation}
\theta_{K_2}(x)=\theta_{K_2^*}(x)=2 \arctan e^{\pm(x-x_0)}
\label{k2kinks} \, \, .
\end{equation}
The energy of the $K_2/K_2^*$ kinks, which also belong to the ${\cal
C}_{\rm NS}$, ${\cal C}_{\rm SN}$ sectors, is greater than the
energy of the $K_1/K_1^*$ kinks: $E^C_{K_2}=E^C_{K_2^*}=2 \lambda
R^2$.


\noindent {\bf 3. Degenerate families of  $Q_\alpha$-kinks.} When
$\sigma^2=1$, the system enjoys $SO(2)$ internal symmetry and the
masses of the two pseudo-Nambu-Goldstone bosons are equal, there are
degenerate families of time-dependent $Q$-kinks of finite energy. If
$\sigma=1$: $ \varphi^{Q_\alpha}(t)=\omega t+\alpha$, where $\omega$
and $\alpha$ are real constants, solves (\ref{pfe2}) for any
time-independent $\theta(x)$. Moreover, by plugging
$\varphi^{Q_\alpha}(t)$ into (\ref{pfe1}) one obtains:
\begin{equation}
\theta^{Q_\alpha}(x)=2\arctan e^{\pm\sqrt{1-\omega^2}(x-x_0)} ,\
\forall\alpha . \label{nlqk}
\end{equation}
Therefore, if $0<\omega^2<1$, the
$(\theta^{Q_\alpha}(x),\varphi^{Q_\alpha}(t))$ configurations form a
degenerate circle of periodic in time Q-kink solutions of energy:
\[
E^C_{Q_\alpha}=\frac{2\lambda R^2}{\sqrt{1-\omega^2}}=
\frac{2\lambda R^2}{\bar{\omega}} \, \, \, \, , \, \, \, \,
\forall\alpha \ .
\]
In fact, these $Q$-kinks can be viewed as the sG kinks rotating
around the main axis of ${\mathbb S}^2$ with constant angular
velocity $\omega$. In another reference frame moving with respect to
the $Q_\alpha$-kink CM with velocity $v$, the interplay between $x$
and $t$ dependence is more complicated:
\begin{eqnarray*}
&&\varphi^{Q_\alpha}(x,t)=\omega\left(\frac{t-vx}{\sqrt{1-v^2}}\right)+\alpha\\
&&\theta^{Q_\alpha}(x,t)=2 \arctan
e^{\pm\bar{\omega}\left(\frac{x-vt}{\sqrt{1-v^2}}-x_0\right)} .
\end{eqnarray*}
At the $\omega=0$ limit we find a circle of static topological kinks
that form a degenerate family of solitary waves of the system.

Of course, all the multi-soliton, soliton-antisoliton and breather
solutions of the sG model are also solitons of our system in the
meridians intersecting either the $\phi_2:\phi_3$ or the
$\phi_1:\phi_3$ planes. We shall not discuss these solutions in this
work and postpone their study to a future research.

\section{Topological kink stability}

\subsection{Small fluctuations on topological kinks}

The analysis of small fluctuations around topological kinks requires
us to consider both the geodesic deviation operator and the Hessian
of the potential energy density. We will denote
$\theta=\theta^1\in[0,\pi]$, $\varphi=\theta^2\in[0,2\pi)$, and thus
the arc-length reads: $ ds^2=R^2 d\theta^1d\theta^1+R^2
\sin^2\theta^1d\theta^2d\theta^2$. We also denote the Kink
trajectories and small deformations around them as:
$\theta_K(x)=(\theta^1_K(x)=\bar{\theta},\theta^2_K(x)=\bar{\varphi})$,
$\theta(x)=\theta_K(x)+\eta(x)$, $\eta(x)=(\eta^1(x),\eta^2(x))$.

Let us consider the following contra-variant vector fields along the
kink trajectory, $\eta , \theta^\prime_K \in \Gamma(T{\mathbb
S}^2\left|_K\right.)$:
$\eta(x)=\eta^1(x)\frac{\partial}{\partial\theta^1}+\eta^2(x)\frac{\partial}{\partial\theta^2}$
and $\theta^\prime_K(x)=\bar{\theta}^\prime
\frac{\partial}{\partial\theta^1}+\bar{\varphi}^\prime\frac{\partial}{\partial\theta^2}$.

The covariant derivative of $\eta(x)$ and the action of the
curvature tensor on $\eta(x)$ are:
\begin{eqnarray*}
&&\nabla_{\theta_K^\prime} \eta=\left(\eta^{\prime
i}(x)+\Gamma^i_{jk}\eta^j\bar{\theta}^{\prime
k}\right)\frac{\partial}{\partial\theta^i}\\ &&
R(\theta^\prime_K,\eta)\theta^\prime_K=\bar{\theta}^{\prime
i}\eta^j(x)\bar{\theta}^{\prime
k}R^l_{ijk}\frac{\partial}{\partial\theta^l}.
\end{eqnarray*}
The geodesic deviation operator is:
\[
\frac{D^2\eta}{dx^2}+
R(\theta^\prime_K,\eta)\theta^\prime_K=\nabla_{\theta_K^\prime}\nabla_{\theta_K^\prime}\eta+
R(\theta^\prime_K,\eta)\theta^\prime_K \   .
\]
To obtain the differential operator that governs the second-order
fluctuations around the kink $\theta_K$, the remaining ingredient is
the Hessian of the potential:
\[
\nabla_\eta {\rm grad}V=\eta^i\left(\frac{\partial^2
V}{\partial\theta^i\partial\theta^j}-\Gamma^k_{ij}\frac{\partial
V}{\partial\theta^k}\right)g^{jl}\frac{\partial}{\partial\theta^l}
\]
evaluated at $\theta_K(x)$. In sum, second-order kink fluctuations
are determined by the operator:
\begin{equation}
\Delta_K\eta=-\left( \nabla_{\theta_K^\prime}\nabla_{\theta_K^\prime}\eta+
R(\theta^\prime_K,\eta)\theta^\prime_K +\nabla_\eta {\rm grad}V\right) \label{eq:hess}
\end{equation}

\subsection{The spectrum of small fluctuations around $K_1/K_1^*$
kinks}

Plugging the $K_1$ solutions into (\ref{eq:hess}), we obtain the
differential operator acting on the second-order fluctuation
operator around the $K_1/K_1^*$ kinks:
\begin{eqnarray}
\Delta_{K_1}\eta&=&\Delta_{K_1^*}\eta=\left[
-\frac{d^2\eta^1}{dx^2}+\left(\sigma^2-\frac{2\sigma^2}{{\rm
cosh}^2\sigma
x}\right)\eta^1\right]\frac{\partial}{\partial\theta^1}\nonumber\\
&& +\left[ -\frac{d^2\eta^2}{dx^2}+2\sigma{\rm tanh}\sigma
x\frac{d\eta^2}{dx}+\bar{\sigma}^2\eta^2\right]\frac{\partial}{\partial\theta^2}\
. \label{eq:hess1}
\end{eqnarray}
The vector fields
$v(x)=v^1(x)\frac{\partial}{\partial\theta^1}+v^2(x)\frac{\partial}{\partial\theta^2}$
parallel along the $K_1$ kink orbits satisfy: $
\frac{dv^i}{dx}+\Gamma^i_{jk}\bar{\theta}^{\prime j} v^k=0$, or,
\[
\left\{\begin{array}{cc} \frac{dv^1}{dx}=0 &
  , \, \, \, v^1(x)=1 \\[0.2cm] \frac{dv^2}{dx}+\sigma\frac{{\rm cotan}(2{\rm
arctan}e^{\sigma x})}{{\rm cosh}\sigma x} v^2=0  & \, \, , \, \, \,
v^2(x)={\rm cosh}\sigma x \end{array}\right. \quad .
\]
Therefore, $v_1=\frac{\partial}{\partial \theta^1}\, \, \, ,\, \, \,
v_2(x)= \cosh \sigma x \, \frac{\partial}{\partial \theta^2}$ is a
frame $\left\{ v_1,v_2\right\}$ in $\Gamma(T{\mathbb S}^2|_{K_1})$
parallel to the $K_1$ kink orbit in which (\ref{eq:hess1}) reads:
\begin{eqnarray}
\Delta_{K_1}\eta&=&\Delta_{K_1^*}\eta=\left[
-\frac{d^2\bar{\eta}^1}{dx^2}+(\sigma^2-\frac{2\sigma^2}{\cosh^2\sigma
x})\bar{\eta}^1\right]\, v_1\nonumber\\ &+& \left[
-\frac{d^2\bar{\eta}^2}{dx^2}+(1-\frac{2\sigma^2}{\cosh^2\sigma
x})\bar{\eta}^2\right]\, v_2\label{eq:hess11}\quad ,
\end{eqnarray}
where $\eta=\bar{\eta}^1 \, v_1+\bar{\eta}^2 \, v_2$,
$\eta^1=\bar{\eta}^1$, and $\eta^2=\cosh \sigma x\, \bar{\eta}^2$.

The second-order fluctuation operator (\ref{eq:hess11}) is a
diagonal matrix of transparent P\"osch-Teller Schr\"odinger
operators with very well known spectra. As expected, despite the
geometric nature of $\bigtriangleup_{K_1}$, we find in the
 $v_1=\frac{\partial}{\partial\theta^1}$ direction the Schr\"odinger
 operator governing sG kink
fluctuations. Finding another P\"osch-Teller potential of the same
type in the $v_2=\frac{\partial}{\partial\theta^2}$ direction  comes
out as a surprise because there is no a priori reason for such a
behavior in the orthogonal direction.

In the $v_1$ direction there is a bound state of zero eigenvalue and
a continuous family of positive eigenfunctions:
\begin{eqnarray*}
\bar{\eta}^1_0(x)=\sech \sigma x &,& \varepsilon^{(1)}_0=0
\\ \bar{\eta}^1_k(x)=e^{i k \sigma x}({\rm tanh}\sigma x - i k) &,&
\varepsilon^{(1)}(k)=\sigma^2(k^2+1) \  .
\end{eqnarray*}
In the $v_2=\cosh\sigma x\frac{\partial}{\partial\theta^2}$
direction the spectrum is similar but the bound state corresponds to
a positive eigenvalue:
\begin{eqnarray*}
\bar{\eta}^2_{1-\sigma^2}(x)=\sech \sigma x
&,& \varepsilon^{(2)}_{1-\sigma^2}=1-\sigma^2>0 \\
\bar{\eta}^2_k(x)=e^{i k \sigma x}({\rm tanh}\sigma x - i k) &,&
\varepsilon^{(2)}(k)=\sigma^2k^2+1 \  .
\end{eqnarray*}
Because there are no fluctuations of negative eigenvalue, the
$K_1/K_1^*$ kinks are stable.

\subsection{One-loop shift to
classical $K_1/K_1^*$ kink masses}

The reflection coefficient of the scattering waves in the potential
wells of the Schr\"odinger operators in (\ref{eq:hess11}) being
zero, it is possible to use the Cahill-Comtet-Glauber formula
\cite{CCG} (see also \cite{BC} for a more detailed derivation) to
compute the quantum correction to the $K_1$ classical kink mass up
to one-loop order:
\begin{eqnarray}
E_{K_1}(\sigma)&=& E^C_{K_1}(\sigma)+\Delta E_{K_1}(\sigma)+{\cal
O}(\frac{1}{R^2})=\nonumber \\&=&2\lambda
R^2\sigma-\frac{\lambda\sigma}{\pi}
[\sin\nu_1+{1\over\sigma}\sin\nu_2-\nu_1\cos\nu_1\nonumber \\&& -
{1\over\sigma}\nu_2\cos\nu_2]+{\cal O}(\frac{1}{R^2})\label{eq:olms}
\end{eqnarray}
In (\ref{eq:olms}) $\nu_1= \arccos 0={\pi\over 2}$, $\nu_2=\arccos
\bar{\sigma}$, are determined from the eigenvalues of the bound
states and the thresholds of the continuous spectra. This simple
structure of the one-loop kink mass shift occurs only for
transparent potentials. In our model, we find the formula:
\begin{equation} E_{K_1}(\sigma) =2\lambda
R^2\sigma-\frac{\lambda\sigma}{\pi}\left[
2-\frac{\bar{\sigma}}{\sigma}\arccos(\bar{\sigma}) \right]+{\cal
O}(\frac{1}{R^2})\label{eq:olms1}
\end{equation}
For instance, for $\sigma={1\over 2}$ we  obtain a result similar to
the mass shift of the $\lambda\phi^4_2$-kink:
\[
E_{K_1}(\frac{1}{2})=\lambda
R^2-\frac{3\lambda}{2\pi}\left(\frac{2}{3}-\frac{\pi}{6\sqrt{3}}\right)+{\cal
O}(\frac{1}{R^2})
\]
As in the $\lambda\phi^4_2$-kink case, a zero mode and a bound
eigenstate of eigenvalue $\varepsilon_{3\over 4}^{(2)}={3\over 4}$
contribute. The gaps between the bound state eigenvalues and the
thresholds $\varepsilon^{(1)}(0)=\sigma^2$, $\varepsilon^{(2)}(0)=1$
of the two branches of the continuous spectrum are the same in our
model. The gaps, however, are different from the gaps in the
$\lambda\phi^4_2$ model between the eigenvalues of the two bound
states and the threshold of the only branch of the continuous
spectrum. Both features contribute to the slightly different result.
The $\sigma=1$ symmetric case is more interesting. We find exactly
twice the spectrum of the sG kink: two zero modes and two gaps with
respect to the thresholds of the continuous spectrum equal to one.
No wonder that the one-loop mass shifts of the degenerate kinks is
twice the one-loop correction of the sG kink:
\[
E_{K_\alpha}(1)=2\lambda \left(R^2-{1\over\pi}\right)+{\cal
O}(\frac{1}{R^2}),\  \forall\alpha \,\,   !\, \, .
\]

Moreover, the quantum fluctuations do not break the $SO(2)$-symmetry
and our result fits in perfectly well with the one-loop shift to the
mass of the ${\cal N}=(2,2)$ SUSY ${\mathbb CP}^1$ kink computed in
\cite{MRNW} where the authors find twice the mass of the ${\cal
N}=1$ SUSY sine-Gordon kink. A different derivation of formula
(\ref{eq:olms1}) following the procedure of \cite{AMAWJ1}, see also
\cite{AMAWJ2, AMAWJ3}, will be published elsewhere.

\subsection{The spectrum of small fluctuations around $K_2/K_2^*$
kinks}

By inserting the $K_2$ solutions in (\ref{eq:hess}) the second-order
fluctuation operator around the $K_2/K_2^*$ kinks is found:
\begin{eqnarray}
&&\Delta_{K_2}\eta=\Delta_{K_2^*}\eta=\left[
-\frac{d^2\eta^1}{dx^2}+(1-\frac{2}{{\rm cosh}^2
x})\eta^1\right]\frac{\partial}{\partial\theta^1}\nonumber\\
&&+\left[ -\frac{d^2\eta^2}{dx^2}+2\tanh
x\frac{d\eta^2}{dx}-\bar{\sigma}^2\eta^2\right]\frac{\partial}{\partial\theta^2}\
. \label{eq:hess2}
\end{eqnarray}
Solving again the parallel transport equations, now along the $K_2$ solutions, it is obtained the parallel frame: $\{u_1,u_2 \}\in\Gamma(T{\mathbb
S}^2\left|_{K_2})\right.$, $u_1= \frac{\partial}{\partial
\theta^1}$, $u_2(x)= \cosh x \, \frac{\partial}{\partial \theta^2}$,
to the $K_2/K_2^*$ orbits. (\ref{eq:hess2}) becomes:
\begin{eqnarray}
\Delta_{K_2}\eta&=&\Delta_{K_2^*}\eta=\left[
-\frac{d^2\tilde{\eta}^1}{dx^2}+(1-\frac{2}{\cosh^2
x})\tilde{\eta}^1\right] u_1\nonumber \\&+&\left[
-\frac{d^2\tilde{\eta}^2}{dx^2}+(\sigma^2-\frac{2}{\cosh^2
x})\tilde{\eta}^2\right] u_2 \label{eq:hess22}\  .
\end{eqnarray}
with $\eta=\tilde{\eta}^1 u_1+\tilde{\eta}^2 u_2$,
$\eta^1=\tilde{\eta}^1$, $\eta^2=\cosh x\tilde{\eta}^2$.

Again, the second-order fluctuation operator (\ref{eq:hess2}) is a
diagonal matrix of transparent P\"osch-Teller operators. In this
case, there is a bound state of zero eigenvalue and a continuous
family of positive eigenfunctions starting at the threshold
$\varepsilon^{(1)}(0)=1$ in the
$u_1=\frac{\partial}{\partial\theta^1}$ direction:
\begin{eqnarray*}
\tilde{\eta}^1_0(x)=\sech  x &,& \varepsilon^{(1)}_0=0 \\
\tilde{\eta}^1_k(x)=e^{i k x}(\tanh x - i k)  &,&
\varepsilon^{(1)}(k)=(k^2+1) \   ,
\end{eqnarray*}
as corresponds to the sG kink. In the $u_2(x)=\cosh
x\frac{\partial}{\partial\theta^2}$ direction, the spectrum is
similar but the eigenvalue of the bound state is negative, whereas
the threshold of this branch of the continuous spectrum is
$\varepsilon^{(2)}(0)=\sigma^2$:
\begin{eqnarray*}
\tilde{\eta}^2_{\sigma^2-1}(x)=\sech x
&,& \varepsilon^{(2)}_{\sigma^2-1}=\sigma^2-1<0 \\
\tilde{\eta}^2_k(x)=e^{i k x}(\tanh x - i k) &,&
\varepsilon^{(2)}(k)=k^2+\sigma^2 \  .
\end{eqnarray*}
Therefore, $K_2/K_2^*$ kinks are unstable and a Jacobi field for
$k=i\sigma$ arises: $\tilde{\eta}^2_J(x)=e^{\sigma x}(\tanh
x-\sigma)$, $\varepsilon^{(2)}_J=0$.

\subsection{One-loop shift to classical $K_2/K_2^*$ kink masses}

Once  again we use the Cahill-Comtet-Glauber formula to compute the
quantum correction to the $K_2$ classical kink mass up to one-loop
order.
As before, the angles $\nu_1= \arccos(0)$ $={\pi\over 2}$, $\nu_2=$
$\arccos(i\bar{\sigma})$, are determined from the eigenvalues of the
bound states and the thresholds of the continuous spectra. The
novelty is that since the bound state eigenvalue is negative $\nu_2$
is purely imaginary. Therefore, we find:
\begin{eqnarray} &&E_{K_2}(\sigma) =2\lambda
R^2-\frac{\lambda\sigma}{\pi}\left[ {1\over\sigma}+\sqrt{2-\sigma^2
}-i{\pi\over 2}\bar{\sigma}\right.\nonumber \\
&&\left.
+\bar{\sigma}\log\left[\sqrt{2-\sigma^2}-\bar{\sigma}\right]\right]+{\cal
O}(\frac{1}{R^2})\  .\label{eq:olms21}
\end{eqnarray}
The key point is that the one-loop mass shift is a complex quantity,
the imaginary part telling us about the life-time of this resonant
state. In the $\sigma=1$ symmetric case, however, we find the
expected purely real answer: $ E_{K_2}(1)=2\lambda
\left(R^2-{1\over\pi}\right)+{\cal O}(\frac{1}{R^2})$.


\subsection{BPS $Q_\alpha$-kinks as $d=1+1$ dyons}

In the $\sigma^2=1$ case there is symmetry with respect to the  $ {\rm
exp}[\alpha{\tiny \left(\begin{array}{ccc} 0 & -1 & 0\\
1 & 0 & 0 \\ 0 & 0 & 0\end{array}\right)}]\in{\mathbb SO}(2)$
subgroup of the ${\mathbb O}(3)$ group. The associated N\"oether
charge distinguishes between different $Q_\alpha$-kinks :
\begin{eqnarray*}
&&Q=\frac{1}{2}\int \, dx \, (\phi_1 \partial_t\phi_2-\phi_2\partial_t\phi_1)=R^2 \int \, dx \,
\sin^2\theta\partial_t\varphi\\
&&Q[Q_\alpha]=R^2\omega\int \, dx \, \sin^2\theta^{Q_\alpha}=2 R^2
\frac{\omega}{\bar{\omega}} \  .
\end{eqnarray*}
For configurations such that $\theta$ is time-independent and
$\varphi$ is space-independent, the energy can be written as:
\begin{eqnarray}
E&=&\frac{\lambda R^2}{2} \int \, dx \,
\left\{\sin^2\theta[\dot{\varphi}-\omega]^2+[\theta'\pm\bar{\omega}\sin\theta]^2\right\}\nonumber
\\ &+& \lambda R^2\int \, dx \, \left\{\omega
\sin^2\theta\dot{\varphi}\mp\bar{\omega}\theta' \sin\theta\right\}
\label{cpns} \ ,
\end{eqnarray}
($\dot{\varphi}=\frac{d\varphi}{dt}(t)$), in such a way that the
solutions of the first-order equations:
\begin{eqnarray*}
&&\dot{\varphi}=\omega \,\Rightarrow \, \varphi^{Q_\alpha}(t)=\omega t+ \alpha \\
&&\theta'=\mp\bar{\omega}\sin \theta \, \Rightarrow \,
\theta^{Q_\alpha}(x)=2\arctan e^{\mp\bar{\omega}(x-x_0)}\ ,
\end{eqnarray*}
the $Q_\alpha$-kinks, saturate the Bogomolny bound and are BPS:
\begin{equation}
E_{\rm BPS}=\frac{2\lambda R^2}{\bar{\omega}}=\lambda\left\{\omega Q
+ \bar{\omega}T\right\} . \label{qbps}
\end{equation}
Here, the topological charge
$T=|W[\theta(+\infty,t)]-W[\theta(-\infty,t)]|$ coming from the
superpotential $W=R^2(1\mp\cos\theta)$ valued at the
$Q_\alpha$-kinks gives: $T[Q_\alpha]=2 R^2, \forall\alpha$. This
explains why \lq\lq one cannot dent a dyon" (even a one-dimensional
cousin), see \cite{CPNS}. Conservation of the N\"oether charge
forbids the decay of $Q_\alpha$ kinks, all of them living in the
same topological sector, to others with less energy.

\subsection{Bohr-Sommerfeld rule: $Q$-kink energy and charge quantization}

The Bohr-Sommerfeld quantization rule applied to periodic in
time-classical solutions in our model reads:
\begin{eqnarray*}
&&\int_0^T \, dt \, \int \, dx \,
\pi_\varphi(t,x)\frac{\partial\varphi}{\partial t}(t,x)\\
&&=R^2 \int_0^T \, dt \, \int \, dx \,
\sin^2\theta(x,t)\frac{\partial\varphi}{\partial
t}\frac{\partial\varphi}{\partial t}=2\pi n \, \, .
\end{eqnarray*}
In \cite{Col1} it is explained how derivation of this formula with
respect to the period $T=\frac{2\pi}{\omega}$ leads to the ODE:
$\lambda\frac{dn}{dE}=\omega^{-1}(E)$, or,
\[
\lambda\int_0^n \, dn=\int_{E_0}^{E_n}\,
\frac{EdE}{\sqrt{E^2-4\lambda^2R^4}}\Rightarrow
E_n=\lambda\sqrt{n^2+4R^4}
\]
starting from $E_0=2\lambda R^2$ and assuming $n$ to be a positive
integer. The $Q$-kink energy is thus quantized and the frequencies
and charges allowed by the Bohr-Sommerfeld rule form also a
numerable infinite set:
\begin{eqnarray*}
\omega_n=\sqrt{1-\frac{1}{1+\frac{n^2}{4 R^4}}} \quad , \quad Q_n=n
\, \, \, \,  .
\end{eqnarray*}

\section{The massive non-linear ${\mathbb S}^2$-sigma model in spherical elliptic coordinates}

The secret of this non-linear (1+1)-dimensional massive ${\mathbb
S}^2$-sigma model is that its analogous mechanical system is
Hamilton-Jacobi separable in spherical elliptic coordinates. This
fact will allow us to known explicitly not only the kink solutions
inherited from the embedded sG models, but the complete set of
solitary waves of the system.

\subsection{The spherical elliptic system of orthogonal coordinates}

The definition of elliptic coordinates in a sphere is as follows:
one fixes two arbitrary points (and the pair of antipodal points) in
${\mathbb S}^2$. We choose these points with no loss of generality
in the form: $ F_1\equiv (\theta_f,\pi)$, $F_2\equiv (\theta_f,0)$,
$\bar{F}_1\equiv (\pi-\theta_f,0)$, $\bar{F}_2\equiv
(\pi-\theta_f,\pi)$, $\theta_f \in (0, {\pi\over 2})$.

The distance between the two fixed points is $d=2f=2R\theta_f<\pi
R$, see Figure 2(a).
\begin{figure}[htbp]
\centerline{\includegraphics[height=3.0cm]{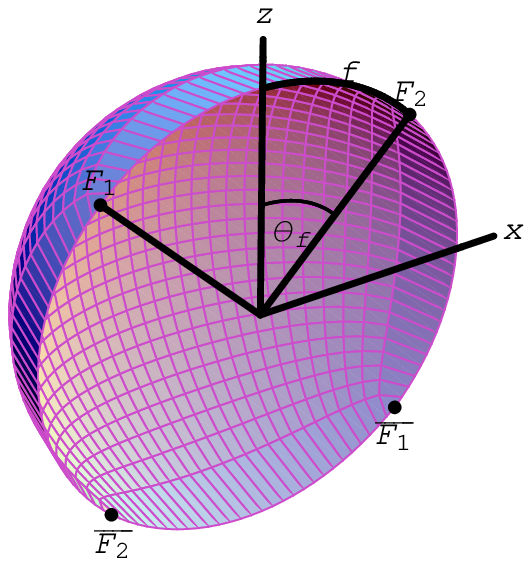}\qquad
\includegraphics[height=2.5cm]{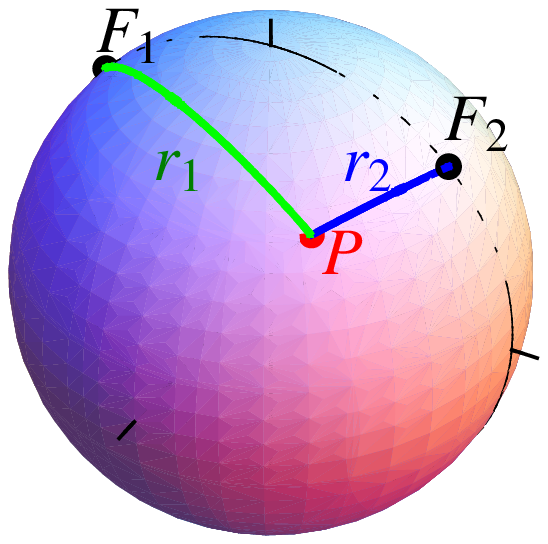}}
{ \caption{a) Foci and antipodal foci of the elliptic system of
coordinates on ${\mathbb S}^2$. b) Distances from a point to the
foci.}}
\end{figure}
Given a point $P\in{\mathbb S}^2$, let us consider the distances
$r_1\in [0,\pi R]$ and $r_2\in [0,\pi R]$ from $P$ to $F_1$ and
$F_2$.
\begin{eqnarray*}
r_1&=& 2R\, \arcsin \sqrt{\frac{1}{2} \left( 1-\cos \theta_f \, \cos
\theta+\sin \theta_f\, \sin \theta \cos\varphi\right)}\\
r_2&=& 2R\, \arcsin \sqrt{\frac{1}{2} \left( 1-\cos \theta_f \, \cos
\theta-\sin \theta_f\, \sin \theta \cos\varphi\right)}\  ,
\end{eqnarray*}
see Figure 2(b). The spherical elliptic coordinates of $P$ are half
the sum and half the difference of $r_1$ and $r_2$: $2u=r_1+r_2$ ,
$2v=r_1-r_2$. $u\in(R\theta_f,R(\pi-\theta_f))$,
$v\in(-R\theta_f,R\theta_f)$. We remark that this version of
elliptic coordinates in a sphere is equivalent to using conical
coordinates constrained to ${\mathbb S}^2$, as defined e. g. in
Reference \cite{MF}. We shall use the abbreviated notation:
\begin{eqnarray*}
\su&=&\sin\frac{u(t,x)}{R},\, \sv=\sin\frac{v(t,x)}{R},\, \sf=\sin\theta_f \\
\su^2&=&\sin^2\frac{u(t,x)}{R} \, , \sv^2=\sin^2\frac{v(t,x)}{R}\, ,
\sf^2=\sin^2\theta_f \
\end{eqnarray*}
and analogously for $\cu$, $\cv$, and $\cf$. To pass from elliptical
to Cartesian coordinates, or viceversa, one uses:
\begin{eqnarray*}
\phi_1(t,x)&=&{R\over\sf}\su \sv\, \, \,  , \quad
\phi_3(t,x)={R\over\cf}\cu \cv
\\ \phi_2(t,x)&=&\pm
\frac{R}{\sf \cf}\sqrt{(\su^2-\sf^2)(\sf^2-\sv^2)}
 \qquad ,
\end{eqnarray*}
whereas the differential arc-length reads:
\[
ds^2_{{\mathbb S}^2}=\frac{\su^2-\sv^2}{\su^2-\sf^2}\cdot
du^2+\frac{\su^2-\sv^2}{\sf^2-\sv^2}\cdot dv^2 \qquad .
\]
The spherical elliptic coordinates of the North and South Poles, and
the foci are respectively: $ (u_N,v_N)=(R\theta_f,0)$,
$(u_S,v_S)=(R(\pi-\theta_f),0)$, $(u_{F_1},v_{F_1})\equiv
(R\theta_f,-R\theta_f)$,  $(u_{F_2},v_{F_2})\equiv
(R\theta_f,R\theta_f)$, $(u_{\bar{F}_1},v_{\bar{F}_1})\equiv
(R(\pi-\theta_f),R\theta_f)$,  $(u_{\bar{F}_2},v_{\bar{F}_2})\equiv
(R(\pi-\theta_f),-R\theta_f)$.

\subsection{Static field equations and Hamilton-Jacobi separability}

We choose a system of spherical elliptic coordinates with the foci
determined by  $\theta_f=\arccos\sigma$, i.e., $\sigma^2={\rm
cos}^2\theta_f$, $\bar\sigma^2={\rm sin}^2\theta_f$. We stress that
the foci (and their antipodal points) are the branching points
mentioned in the previous Section. In this coordinate system the
action for the massive non-linear ${\mathbb S}^2$-sigma model reads:
\begin{eqnarray*}
&&S=\int dtdx
\left\{\frac{1}{2}\left[\frac{\su^2-\sv^2}{\su^2-\sf^2}\partial_\mu
u\partial^\mu u+\frac{\su^2-\sv^2}{\sf^2-\sv^2}\partial_\mu
v\partial^\mu v\right]\right. \\
&&\hspace{2cm} \left.-V(u(t,x),v(t,x))
\rule{0cm}{0.4cm}\right\}\  ,\\
&&V(u,v) = \frac{R^2}{2(\su^2 -\sv^2)} \left[ \su^2 (\su^2-\sf^2) +
\sv^2 (\sf^2 -\sv^2) \right] \, .
\end{eqnarray*}
The static energy reads:
\begin{eqnarray*}
&&E[u,v]=\lambda \int dx  \, {\cal E}(u^\prime(x), v^\prime(x), u(x), v(x))\   ,\\
&& {\cal E}=\frac{1}{2}\left[\frac{\su^2-\sv^2}{\su^2-\sf^2}(u')^2+
\frac{\su^2-\sv^2}{\sf^2-\sv^2}(v')^2\right] +V(u,v) \, \, .
\end{eqnarray*}
Let us think of $E[u,v]$ as the action for a particle: ${\cal E}$ as
the Lagrangian, $x$ as the time, $U(u,v)=-V(u,v)$ as the mechanical
potential energy, and the target manifold ${\mathbb S}^2$ as the
configuration space. The canonical momenta are:
$p_u=\frac{\partial{\cal E}}{\partial u^\prime}$,
$p_v=\frac{\partial{\cal E}}{\partial v^\prime}$, and the static
field equations can be thought of as the Newtonian ODE's:
\begin{eqnarray*}
&& \frac{d}{d x}\cdot\left(\frac{\su^2-\sv^2}{\su^2-\sf^2}\cdot u'
\right)= \frac{\delta V}{\delta u}\\ && \frac{d}{d
x}\cdot\left(\frac{\su^2-\sv^2}{\sf^2-\sv^2}\cdot v'\right)=
\frac{\delta V}{\delta v} \qquad .
\end{eqnarray*}
Because the mechanical energy is
\begin{eqnarray*}
U(u,v)&=&-V(u,v)=-\frac{1}{\su^2-\sv^2} \, \left(f(u)+g(v)\right)=\\
&=& -\frac{R^2[\su^2(\su^2-\sf^2)+\sv
^2(\sf^2-\sv^2)]}{2(\su^2-\sv^2)}
\end{eqnarray*}
this mechanical system is a Liouville type I integrable system, (see
\cite{Per}). The Hamiltonian and the Hamilton-Jacobi equation of
spherical Type I Liouville models have the form:
\begin{eqnarray*}
&&H=\frac{h_u+h_v}{\su^2-\sv^2} \, , \   \left\{
\begin{array}{c} h_u=\frac{1}{2} (\su^2-\sf^2)\, p_u^2-f(u)\\ h_v=\frac{1}{2} (\sf^2-\sv^2)\,
p_v^2-g(v) \end{array} \right. \\ &&\frac{\partial {\cal
S}}{\partial x} +H\left(\frac{\partial {\cal S}}{\partial
u},\frac{\partial {\cal S}}{\partial v},u,v\right)=0 \ ,
\end{eqnarray*}
which guarantees HJ separability in this system of coordinates. The
separation ansatz ${\cal S}(x,u,v)$ $=-i_1 x+ {\cal S}_u(u)+{\cal
S}_v(v)$ reduces the HJ equation to the two separated ODE's, in the
usual HJ procedure, leading to the complete solution:
${\cal S}={\cal S}(x,u,v,i_1,i_2)$:
\begin{eqnarray}
{\cal S}&=& -i_1 x+{\rm sg}(p_u) \int du
\sqrt{\frac{2(\frac{i_2}{R^2}+i_1 \su^2+f(u))}{\su^2-\sf^2}}\nonumber\\
&&+{\rm sg}(p_v) \int dv \sqrt{\frac{2(-\frac{i_2}{R^2}-i_1
\sv^2+g(v))}{\sf^2-\sv^2}} \label{eq:pham}
\end{eqnarray}
in terms of the mechanical energy $I_1=i_1$ and a second constant of
motion: the separation constant $I_2=\frac{i_2}{R^2}$.

\section{Non-topological kinks}

We now identify the families of trajectories corresponding to the
values $i_1=i_2=0$ of the two invariants in the mechanical system.
These orbits are separatrices between bounded and unbounded motion
in phase space and become solitary wave solutions in the
field-theoretical model because the $i_1=i_2=0$ conditions force the
boundary behavior (\ref{febc}). (See \cite{Ito} and \cite{AAi} for
application of this idea to the search for solitary waves in other
two-scalar field models with analogous mechanical systems which are
HJ separable in elliptic coordinates.)

\noindent 1. In a first step we find the Hamilton characteristic
function for zero particle energy $(i_1=0=i_2)$ by performing the
integrations in (\ref{eq:pham}): $W(u,v)={\cal
S}_u(u,i_1=0,i_2=0)+{\cal S}_v(v,i_1=0,i_2=0)$,
\[
W^{(\beta_1,\beta_2)}(u,v)=F^{(\beta_1)}(u)+G^{(\beta_2)}(v)
\]
with $(-1)^{\beta_1}=-{\rm sg}\,p_u$,  $ (-1)^{\beta_2} =-{{\rm
sg}\,p_v}\cdot {\rm sg}\,v$, and:
\[
F^{(\beta_1)}(u)= R^2(-1)^{\beta_1} \cu\, ,\quad
G^{(\beta_2)}(v)=R^2(-1)^{\beta_2} \cv\, .
\]

\noindent 2. The HJ procedure provides the kink orbits by
integrating ${\rm sg}\, p_u \int \frac{du}{(\su^2-\sf^2)|\su|} -
{\rm sg}\, p_v \int \frac{dv}{(\sf^2-\sv^2)|\sv|} = R^3 \gamma_2$:
\begin{eqnarray}
e^{R^2 \gamma_2 \sf^2} &=& \left[ \frac{\left|\tan \frac{u-f}{2R}
\tan \frac{u+f}{2R}\right|^{\frac{1}{2\cf }}}{|\tan \frac{u}{2R}|}
\right]^{{\rm sg}p_u} \cdot \nonumber\\ && \left[ \frac{|\tan
\frac{v}{2R}|}{\left|\tan \frac{v-f}{2R} \tan
\frac{v+f}{2R}\right|^{\frac{1}{2\cf }}} \right]^{{\rm sg}p_v}
\label{eq:ntkor} \  .
\end{eqnarray}
In Figure 3(a) a Mathematica plot is offered showing several orbits
complying with (\ref{eq:ntkor}) for several values of the
integration constant $\gamma_2$. Note that all the orbits start and
end at the North Pole and pass through the foci $\bar{F}_1$ such
that we have shown a one-parametric family of non-topological kink
orbits. In fact, there are four families of non-topological kinks
among the solutions of (\ref{eq:ntkor}): the orbits of a second
family also start and end at the North Pole but pass through
$\bar{F}_2$. The orbits in the second pair of NTK families start and
end at the South Pole ant pass through either $F_1$ or $F_2$.
\begin{figure}[h]
\centerline{\includegraphics[height=3.0cm]{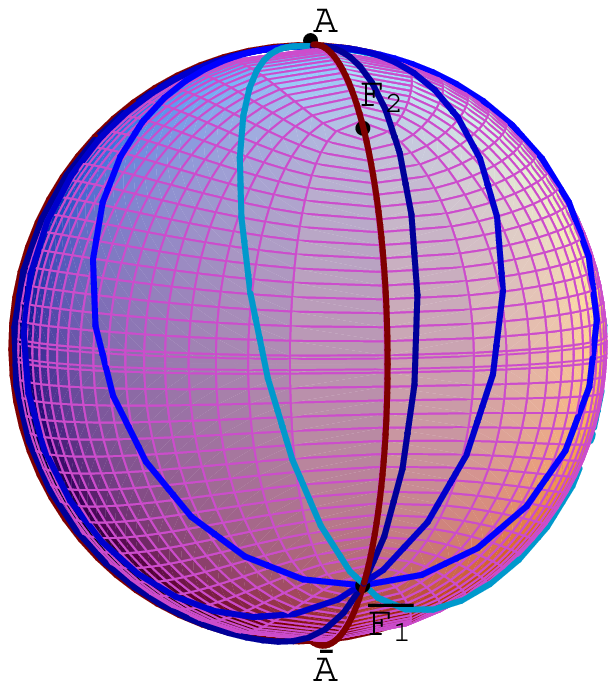}\quad
\includegraphics[height=3.0cm]{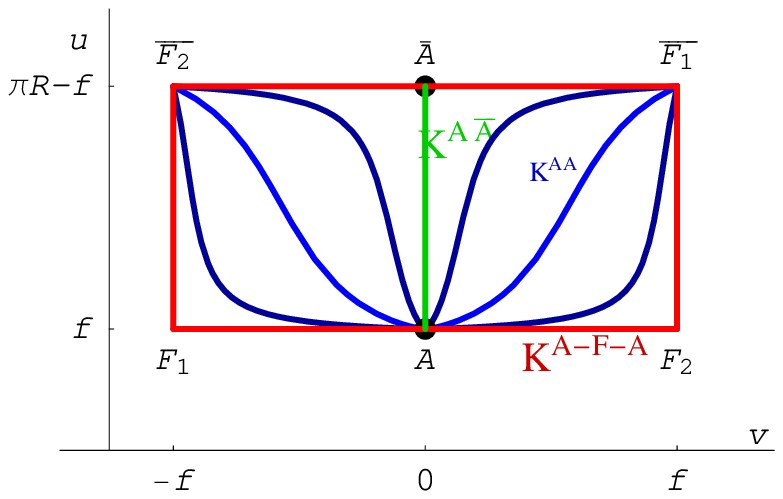}}
\caption{a) Several NTK kink orbits. b) The same NTK kink orbits in
the elliptic rectangle.}
\end{figure}
\begin{figure}[htbp]
\centerline{\includegraphics[height=2.5cm]{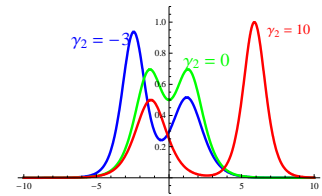}}
\caption{NTK energy densities for three different values of
$\gamma_2$: 1) $\gamma_2=-3$, highest peak on the left (blue) 2)
$\gamma_2=0$, symmetrical peaks (green) 3) $\gamma_2=10$ highest
peak on the right (red).}
\end{figure}

\noindent 3. The HJ procedure requires similar integrations in ${\rm
sg}\, p_u \int \frac{|\su| du}{(\su^2-\sf^2)} - {\rm sg}\, p_v \int
\frac{|\sv| dv}{(\sf^2-\sv^2)} = R(x+ \gamma_1)$ to find the kink
profiles (or particle \lq\lq time " schedules):
\begin{equation}
e^{2(x+\gamma_1) \cf} =  \frac{\left|\tan \frac{u(x)-f}{2R} \tan
\frac{u(x)+f}{2R}\right|^{{\rm sg}p_u}}{\left|\tan \frac{v(x)-f}{2R}
\tan \frac{v(x)+f}{2R}\right|^{{\rm sg}p_v}} \label{eq:ntkprof}
\qquad .
\end{equation}
\noindent In Figure 4 the NTK energy densities for three values of
$\gamma_2$ are plotted.

\noindent 4. Reshuffling equations (\ref{eq:ntkor}) and
(\ref{eq:ntkprof}), it is possible to find the NTK families
analytically, (\ref{eq:ntksol}), based on $(u_N,v_N)=(R\theta_f,0)$. The other
families, based on $(u_S,v_S)=(R(\pi-\theta_f),0)$ are given by a
similar formula.

\begin{widetext}
{\small
\begin{eqnarray}
\tan{u_K(x,\gamma_1,\gamma_2)\over 2 R} & =& \frac{\pm \sqrt{2}
\sqrt{1+e_1} e_2 \tfmedios}{\sqrt{e_1+e_2^2 + \tfmedios^4 +e_1 e_2^2
\tfmedios^4 - \sqrt{(e_1+e_2^2+\tfmedios^4 +e_1 e_2^2 \tfmedios^4)^2
- 4(1+e_1)^2 e_2^2 \tfmedios^4}}} \nonumber \\
\tan{v_K(x,\gamma_1,\gamma_2)\over 2 R} & =&
\frac{\pm\sqrt{e_1+e_2^2 + \tfmedios^4 +e_1 e_2^2 \tfmedios^4 -
\sqrt{(e_1+e_2^2+\tfmedios^4 +e_1 e_2^2 \tfmedios^4)^2 - 4(1+e_1)^2
e_2^2 \tfmedios^4}}}{\sqrt{2} \sqrt{1+e_1} \tfmedios}
\label{eq:ntksol}
\end{eqnarray}
}
where we have used the new abbreviations: $e_1=e^{2(x+\gamma_1)\cf}
\, \, \,  , \, \, \, e_2=e^{x+\gamma_1-R^2 \gamma_2 \sf^2} \, \, \,
, \, \, \, \tfmedios=\tan\frac{f}{2R}$.
\end{widetext}

\section{Non-topological kink instability: Morse index theorem}

To study the (lack of) stability of NTK kinks, it is convenient to
use the following notation for the elliptic variables: $u^1=u$,
$u^2=v$. The static field equations read:
\begin{equation}
\frac{D}{dx}\cdot\frac{du^i}{dx}=g^{ij}\frac{d}{dx}\left(g_{jk}\frac{du^k}{dx}\right)=g^{ij}\frac{\partial
V}{\partial u^j} \ . \label{ssoe}
\end{equation}
Let us consider a one-parametric family of solutions of
(\ref{ssoe}): $u^i_K(x;\gamma)$. The derivation of
\[
\left(-\frac{D}{dx}\cdot\frac{du^i_K}{dx}+g^{ij}(u^1_K,u^2_K)\frac{\partial
V}{\partial u^j}\right)\cdot g_{ik}\frac{\partial
u^k_K}{\partial\gamma}=0
\]
with respect to the parameter $\gamma$ implies:
\begin{eqnarray*}
&&\frac{D^2}{dx^2}\cdot\frac{\partial
u^i_K}{\partial\gamma}+\frac{\partial u^j_K}{\partial
x}\cdot\frac{\partial u^k_K}{\partial\gamma}\cdot\frac{\partial
u^l_K}{\partial x}R^i_{jkl}+ \\
&&+g^{ik}\left(\frac{\partial^2 V}{\partial u^j\partial
u^k}-\Gamma^l_{jk}(u^1_K,u^2_K)\frac{\partial V}{\partial
u^l}\right)\frac{\partial u^j_K}{\partial\gamma}=0\, .
\end{eqnarray*}
In the last three formulas the metric tensor, the covariant
derivatives, the connection, the curvature tensor, and the gradient
and Hessian of the potential are valued on $(u^1_K,u^2_K)$, see
\cite{AMAJ2}. Thus, $\frac{\partial u^i_K}{\partial\gamma}$ is an
eigenvector of the second order fluctuation operator of zero
eigenvalue. The derivatives of the NTK solutions (\ref{eq:ntksol})
with respect to the parameter $\gamma_2$ are accordingly
eigenvectors of the second order fluctuation operator of zero
eigenvalues orthogonal to the NTK orbit, i.e., Jacobi fields that
move from one NTK kink to another with no cost in energy.

Better than direct derivation of (\ref{eq:ntksol}) the Jabobi fields
can be obtained from (\ref{eq:ntkor}) and (\ref{eq:ntkprof}) by
using implicit derivation with respect to the parameter $\gamma_2$
and solving the subsequent linear system. We skip the (deep)
subtleties of this calculation and merely provide the explicit
analytical expressions:
\begin{eqnarray}
&& J^{\rm NTK}(x;\gamma_2)= \frac{R^3 (\su^2-\sf^2)
(\sf^2-\sv^2)}{\su^2-\sv^2} \cdot \nonumber\\ && \hspace{1cm}
\cdot\left( {\rm sg}(p_u) \, \su \, \frac{\partial }{\partial u} \,
-\, {\rm sg}(p_v) \, \sv \, \frac{\partial }{\partial v} \right) \
.\label{eq:jacf}
\end{eqnarray}
In Figures 5 a)-b), 6 a)-b) two Jacobi fields for two values of
$\gamma_2$, as well as the corresponding NTK field profiles, are
plotted for the three $\phi_1$, $\phi_2$, $\phi_3$ original field
components.
\begin{figure}[htbp]
\centerline{ {\includegraphics[height=2.5cm]{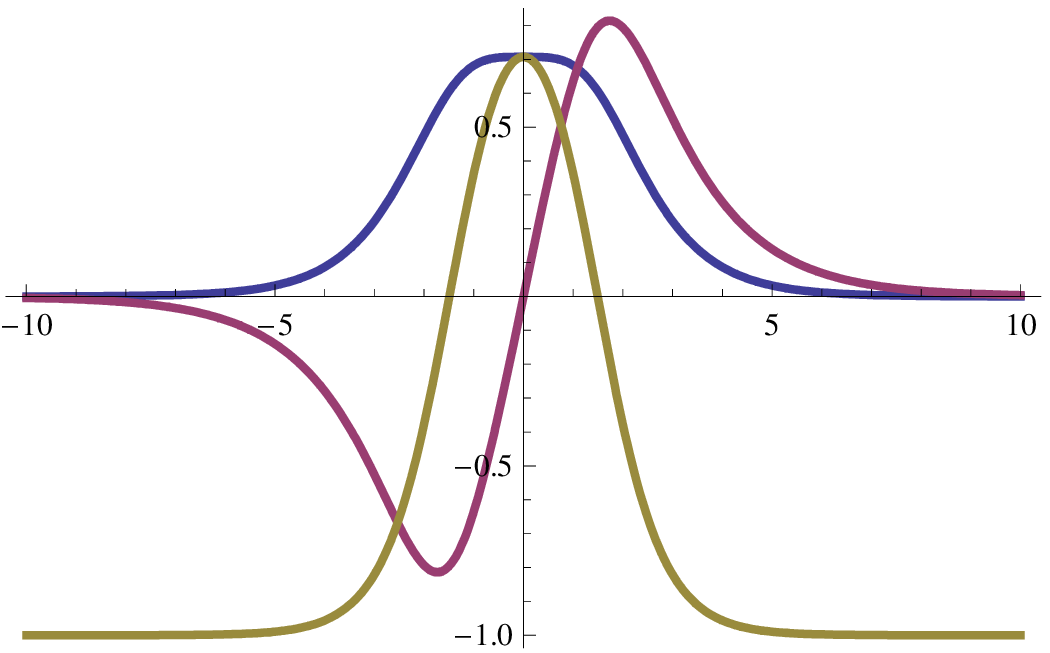}} \
\includegraphics[height=2.5cm]{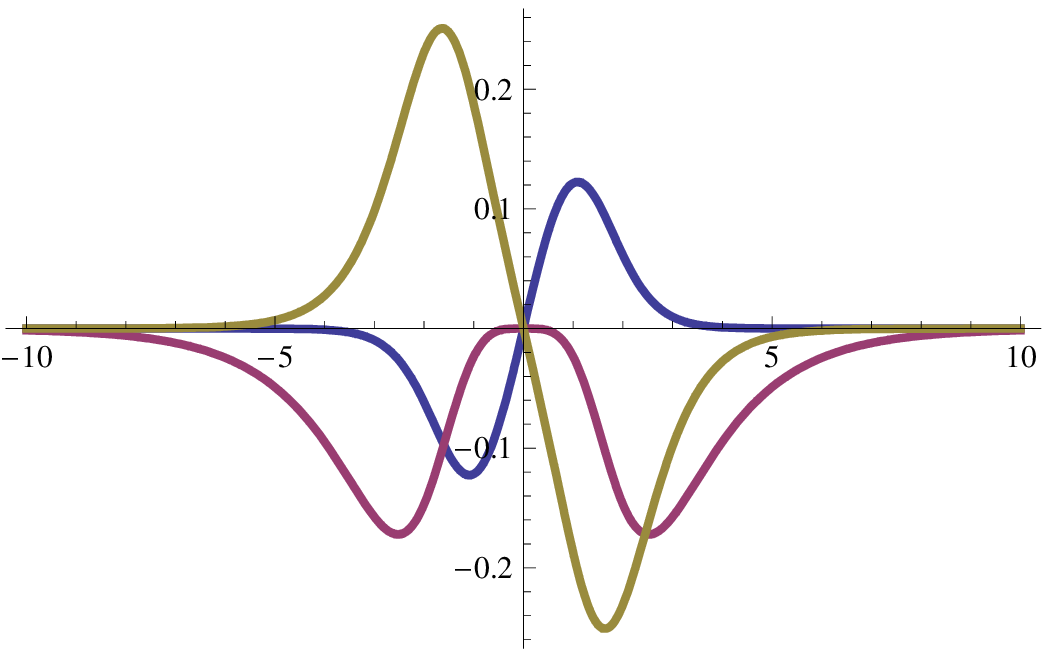}}
\caption{ a) Profiles of the field components for NTK $\gamma_2=0$
kink. b) Plot of the Jacobi field $J^{NTK}(x;0)$}
\end{figure}
\begin{figure}[htbp]
\centerline{ {\includegraphics[height=2.5cm]{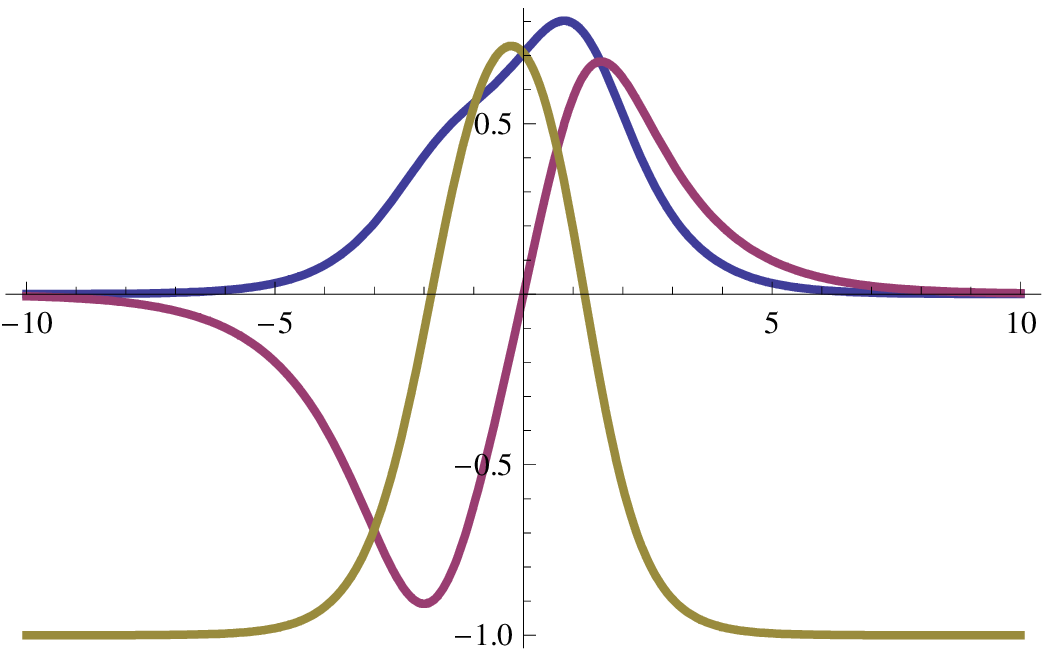}} \
\includegraphics[height=2.5cm]{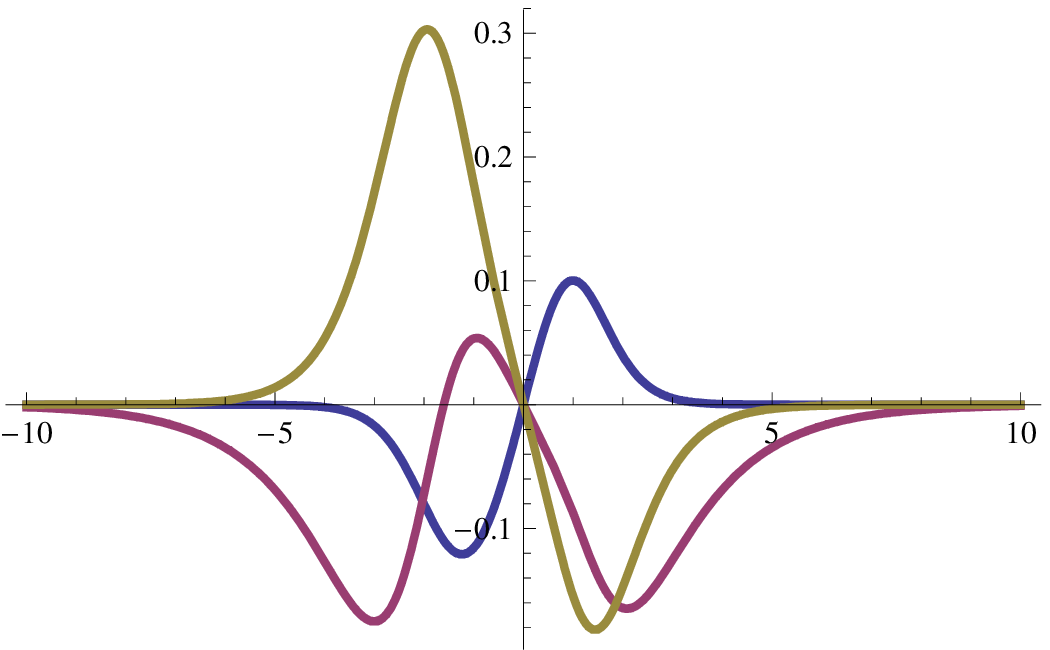}}
\caption{ a) Profiles of the field components for NTK $\gamma_2=1$
kink. b) Plot of the Jacobi field $J^{NTK}(x;1)$}
\end{figure}

The zeroes of the Jacobi fields along a given $\gamma_2$-NTK orbit
(in the four disconnected sectors) are as follows: either $A\equiv
(u_K(-\infty;\gamma_2)=f, v_k(-\infty;\gamma_2)=0$),
$\bar{F}_1\equiv (u_K(\bar{\gamma}_1;\gamma_2)=\pi R-f,
v_K(\bar{\gamma}_1;\gamma_2)=-f$), $\bar{F}_2\equiv
(u_K(\bar{\gamma}_1;\gamma_2)=\pi R-f,
v_K(\bar{\gamma}_1;\gamma_2)=f$), or, $\bar{A}\equiv
(u_K(-\infty;\gamma_2)=\pi R-f, v_k(-\infty;\gamma_2)=0$),
$F_1\equiv (u_K(\bar{\gamma}_1;\gamma_2)=f,
v_K(\bar{\gamma}_1;\gamma_2)=-f$), $F_2\equiv
(u_K(\bar{\gamma}_1;\gamma_2)=f, v_K(\bar{\gamma}_1;\gamma_2)=f$).
Thus, the conjugate points with respect to either the North or the
South Poles along the NTK orbits are listed below: {\small
\[
\begin{array}{|c|c|c|} \hline
& &  \\[-0.1cm]
{\bf  Starting} \, {\bf Point}\, & {\bf Conjugate}\, {\bf Point} & {\bf Conjugate}\, {\bf Point} \\ \hline &&  \\
  [-0.1cm]  {\rm North}\, {\rm Pole}:  A
  &  {\rm Antipodal}\, {\rm Focus}:   \bar{F}_1 &
  {\rm Antipodal}\, {\rm Focus}:   \bar{F}_2
 \\ \hline &&  \\[-0.1cm]
  {\rm South}\, {\rm Pole}:  \bar{A} &
{\rm Focus}:   F_1 &
{\rm Focus}:    F_2 \\
\hline
\end{array}
\]}
In this two-dimensional setting, the Morse index theorem states that
the number of negative eigenvalues of the second order fluctuation
operator around a given orbit is equal to the number of conjugate
points crossed by the orbit \cite{ItoT}. The reason is that the
spectrum of the Schr$\ddot{\rm o}$dinger operator has in this case
an eigenfunction with as many nodes as the Morse index, the Jacobi
field, whereas the ground state has no nodes. The Jacobi fields of
the NTK orbits cross one conjugate point, their Morse index is one,
and the NTK kinks are unstable.

\section{Non-BPS non-topological kinks} The availability of the
Hamilton characteristic function as a sum of one function of $u$ and
another function of $v$ allows us to write the energy of static
configurations á la Bogomolny:
\begin{eqnarray*}
E[u,v]&=&{\lambda\over 2}\int dx
\left\{\frac{\su^2-\sv^2}{\su^2-\sf^2}\left(\frac{du}{dx}-
\frac{\su^2-\sf^2}{\su^2-\su^2}\frac{dF^{(\beta_1)}}{du}\right)^2\right.\\
 &+&\left. \frac{\su^2-\sv^2}{\sf^2-\sv^2}\left(\frac{dv}{dx}-
\frac{\sf^2-\sv^2}{\su^2-\su^2}\frac{dG^{(\beta_2)}}{dv}\right)^2 \right\}\\
&+&\lambda \int \, dx \,
\frac{du}{dx}\frac{dF^{(\beta_1)}}{du}+\lambda \int \, dx \,
\frac{dv}{dx}\frac{dG^{(\beta_2)}}{dv} \  .
\end{eqnarray*}

Solutions of the first-order equations
\begin{eqnarray}
\frac{du}{dx}&\!=\!&\frac{\su^2-\sf^2}{\su^2-\sv^2}\frac{dF^{(\beta_1)}}{du}
=-R(-1)^{\beta_1}\frac{\su^2-\sf^2}{\su^2-\sv^2}\su \label{eq:efoe}\\
\frac{dv}{dx}&\!=\!&\frac{\sf^2-\sv^2}{\su^2-\sv^2}\frac{dG^{(\beta_2)}}{dv}
=-R(-1)^{\beta_2}\frac{\sf^2-\sv^2}{\su^2-\sv^2}\sv \label{eq:efoe1}
\end{eqnarray}
are absolute minima of the energy and therefore are stable. Note
that the energy of the solutions of (\ref{eq:efoe})-(\ref{eq:efoe1})
is positive or zero because ${\rm sg}u^\prime={\rm
sg}\frac{dF^{(\beta_1)}}{du}$ and ${\rm sg}v^\prime={\rm
sg}\frac{dG^{(\beta_1)}}{dv}$.

Even though the NTK trajectories are solutions of the analogous
mechanical system provided by the HJ procedure that is closely
related to the ODE system (\ref{eq:efoe})-(\ref{eq:efoe1}), they do
not strictly solve (\ref{eq:efoe})-(\ref{eq:efoe1}). Taking the
quotient of the two equations in (\ref{eq:efoe})-(\ref{eq:efoe1}) we
find the equation
\begin{equation}
\frac{du}{dv}=(-1)^{\beta_1-\beta_2}\frac{\su^2-\sf^2}{\sf^2-\sv^2}\,\frac{\su}{\sv}
\qquad , \label{eq:efokf}
\end{equation}
which determines the kink orbit flow. Note that this equation is
identical to the equation in the HJ procedure that one must
integrate to find (\ref{eq:ntkor}). The subtle point, however, is
that this flow is undefined, $\frac{0}{0}$, at the four foci: $F_1$,
$F_2$, $\bar{F}_1$, $\bar{F}_2$, and all the NTK orbits pass through
one of these dangerous points, see Figures 3(a) and 3(b). The
non-topological kink orbits solve (\ref{eq:efoe})-(\ref{eq:efoe1})
for a given sign combination before meeting at a focus and are
solutions of (\ref{eq:efoe})-(\ref{eq:efoe1}) with another choice of
signs after leaving these orbit intersections. Thus, non-topological
kinks are classified as non-BPS in the terminology of \lq\lq
pre-supersymmetric" systems. We remark that in elliptic coordinates
the pathology is not in the Hamilton characteristic function but in
the factors induced by the change to elliptic coordinates. The
conclusion is that the energy of the NTK kinks must be computed
piecewise along the orbit. $E_{K(\gamma_2)}^C=2\lambda
\left|G^{(\beta_2)}(0)-G^{(\beta_2)}(v^\pm_B)\right|+ 2\lambda
\left|F^{(\beta_1)}(u^+_B)-F^{(\beta_1)}(u^-_B)\right|$, i.e.,
\begin{equation}
E_{K(\gamma_2)}^C=2\lambda R^2|1-\sigma|+2\lambda R^2|2\sigma|
=2\lambda R^2(1+\sigma) \label{eq:ntken}
\end{equation}
gives the kink energy as the action of the corresponding trajectory.

\subsection{Singular $K_1$ and $K_2$ kinks: kink
mass sum rule}

Analysis of the BPS/non-BPS nature of the topological kinks in
elliptic coordinates is illuminating. The $K_1/K_1^*$ kink orbits
lie in the $v=0$ line, splitting the two-halves of the elliptic
rectangle: $v_{K_1}=v_{K_1^*}=0$, see Figure 3(b). The first-order
equations (\ref{eq:efoe})-(\ref{eq:efoe1}) on the $K_1/K_1^*$ kink
orbits ($\beta_1=0$ gives kinks and $\beta_1=1$ anti-kinks) and the
$K_1/K_1^*$ kink profiles in elliptic coordinates are:
\begin{eqnarray*}
&&\frac{du}{dx}=-(-1)^{\beta_1}R\frac{\su^2-\sf^2}{\su}\\&&
u_{K_1}(x)=u_{K_1^*}(x)=R \arccos [\sigma\tanh((-1)^{\beta_1}\sigma
x)] \ .
\end{eqnarray*}
The $K_1/K_1^*$ kink energy saturates the BPS bound:
\[
E_{K_1}^C=\lambda
\left|F^{(\beta_1)}(u_{K_1}(+\infty))-F^{(\beta_1)}(u_{K_1}(-\infty))\right|=2\lambda
R^2\sigma  .
\]

The $K_2/K_2^*$ kink orbits are the four edges of the elliptic
rectangle: $u_{K_2}=u_{K_2^*}=R\theta_f$, $v_{K_2}=R\theta_f$,
$v_{K_2}=-R\theta_f$, $u_{K_2}=u_{K_2^*}=R(\pi-\theta_f)$, see again
Figure 3(b). The $K_2/K_2^*$ kinks are accordingly three-step
trajectories in the elliptic rectangle.

I.  $-\infty < x< \log\tan{\theta_f\over 2}$ and
$u_{K_2}^I=u_{K_2^*}^I=R(\pi-\theta_f)$, the first-order ODE, and
the solutions are:
\[
\beta_2=1,\,  v'=R|\sv| ,\, v^I_{K_2}(x)=-v^I_{K_2^*}(x)=2R\arctan
e^x\, .
\]

II.  $\log\tan{\theta_f\over 2}< x< \log\tan{\pi-\theta_f\over 2}$,
$v_{K_2}^{II}=-v_{K_2^*}^{II}=R\theta_f$, the first-order ODE and
the solution are:
\[
\beta_1=0,\, u'=-R\su,\, u^{II}_{K_2}(x)=u^{II}_{K_2^*}(x)=2R\arctan
e^{-x}\, .
\]

III. $\log\tan{\pi-\theta_f\over 2}< x < +\infty $,
$u_{K_2}^{III}=u_{K_2^*}^{III}$, the first-order equation and the
solutions are:
\[
\beta_2=0,\, v'=-R|\sv|
,\,v^{III}_{K_2}(x)=-v^{III}_{K_2^*}(x)=2R\arctan e^{-x}\, .
\]
Anti-kinks are obtained by changing the choices of $\beta_1$ and
$\beta_2$. In any case, the $K_2/K_2^*$ kink energy is not of the
BPS form:
\begin{eqnarray*}
&&E_{K_2}^C=\lambda
\left|G^{(1)}(v(-\infty))-G^{(1)}(v(\log\tan\frac{\theta_f}{2}))\right|\\
&&+ \lambda \left|F^{(0)}(u(\log{\rm
tan}\frac{\theta_f}{2}))-F^{(0)}(u(\log{\rm
cotan}\frac{\theta_f}{2}))\right|\\&&+\lambda
\left|G^{(0)}(v(\log{\rm
cotan}\frac{\theta_f}{2}))-G^{(0)}(v(+\infty))\right|\\ &&  =
\lambda R^2|1-\cf|+\lambda R^2|-2\cf|+\lambda R^2|1-\cf+1|=2\lambda
R^2 .
\end{eqnarray*}
It is remarkable that these energies satisfy the following \lq\lq
Kink mass sum rule":
\begin{equation}
E_{K(\gamma_2)}^C=2\lambda R^2(1+\sigma)=E_{K_2}^C+E_{K_1}^C
\label{kmsr}
\end{equation}
In fact, the $|\gamma_2|\to\infty$ limit of the family of
$K_{\gamma_2}$ (NTK) kinks is compatible with equation
(\ref{eq:ntkor}) only at the edges of the elliptic rectangle
(forming the $K_2/K_2^*$ orbits) and the $K_1/K_1^*$ orbit.
Therefore, the $K_1$ and $K_2$ form the boundary of the moduli space
of $K_{\gamma_2}$ in such a way that (\ref{kmsr}) shows this
combination as one of the NTK kinks.

\section{Solitary spin waves}

Field configurations that satisfy the Euler-Lagrange equations:
\begin{eqnarray*}
&&\frac{\partial A_a}{\partial t}(t,x)=\sum_{b=1}^3\,
\left(\frac{\delta A_b}{\delta\phi_a}(t,x)-\frac{\delta
A_a}{\delta\phi_b}(t,x)\right)\cdot\frac{\partial\phi_b}{\partial
t}(t,x)\\ && =\sum_{b=1}^3\,\sum_{c=1}^3 \,
\varepsilon_{abc}B_c[\Phi(t,x)]\cdot\frac{\partial\phi_b}{\partial
t}(t,x)
\end{eqnarray*}
are extremals of the \lq\lq Wess-Zumino" action:
\[
S_{\rm WZ}[\Phi]=R^2\int \, dtdx \, \sum_{a=1}^3\,
A_a[\Phi(t,x)]\frac{\partial\phi_a}{\partial t}(t,x) \qquad \qquad .
\]
In particular a \lq\lq magnetic monopole " field
$B_a[\Phi(t,x)]=\frac{\phi_a(t,x)}{R^3}$ in the ${\mathbb R}^3$
internal space where the ${\mathbb S}^2$-sphere is embedded is
obtained by the choice of singular \lq\lq vector potentials":
\begin{eqnarray*}
A_1^\pm[\Phi(t,x)]&=&-\frac{\phi_2}{\sqrt{\phi_1^2+\phi_2^2+\phi_3^2}(\phi_3\pm
\sqrt{\phi_1^2+\phi_2^2+\phi_3^2})} \\
A_2^\pm[\Phi(t,x)]&=&\frac{\phi_1}{\sqrt{\phi_1^2+\phi_2^2+\phi_3^2}(\phi_3\pm
\sqrt{\phi_1^2+\phi_2^2+\phi_3^2})} \\ A_3^\pm[\Phi(t,x)]&=&0 \qquad
\qquad .
\end{eqnarray*}
$\vec{A}^+[\Phi(t,x)]]$ is singular on the negative $\phi_3$-axis
but a gauge transformation to $\vec{A}^-[\Phi(t,x)]]$ moves the
Dirac string -henceforth a gauge artifact- to the positive
$\phi_3$-axis. The scalar fields are constrained to live in the
$\phi_1^2(t,x)+\phi_2^2(t,x)+\phi_3^2(t,x)=R^2$ sphere, a surface
where this magnetic flux is constant. Therefore, Stoke's theorem
tells us that $S_{\rm WZ}=R^2\int \, dx \, \oint\, \sum_{a=1}^3
d\phi_a(x)A_a[\Phi(x)]$ is the area bounded by a closed curve in
${\mathbb S}^2$.

The important point is that the Euler-Lagrange equations for the sum
of the two actions $S_{\rm WZ}+S$, where $S$ is the action of our
model, are:
\begin{equation}
{1\over R}\sum_{b=1}^3\, \sum_{c=1}^3 \,
\varepsilon_{abc}\phi_c\frac{\partial\phi_b}{\partial
t}+\Box\phi_a+\frac{\alpha_a^2}{\lambda^2}\phi_a=0 \label{llf} \  .
\end{equation}
At the long wavelength limit, the ODE system ({\ref{llf}) become the
Landau-Lifshitz system of equations of ferromagnetism. The
connection between the semi-classical (high-spin) limit of the
Heisenberg model and the quantum non-linear ${\mathbb S}^2$-sigma
model is well established \cite{Hal}.

\subsection{Spin waves}

Plugging the constraint into ({\ref{llf}), we find the system of two
ODE's:
\begin{eqnarray}
&-&\sum_\alpha \varepsilon_{\alpha\beta}\frac{{\rm
sg}\phi_3}{R}\left(\sqrt{R^2-\sum_\gamma\phi_\gamma\phi_\gamma}\cdot\frac{\partial\phi_\alpha}{\partial
t}\right.\nonumber
\\ && \left.+\phi_\alpha\frac{\sum_\gamma\phi_\gamma\partial_t\phi_\gamma}{\sqrt{R^2-\sum_\gamma\phi_\gamma\phi_\gamma}}
\right) +\Box\phi_\beta + m_\beta^2\phi_\beta\nonumber
\\&+&\frac{\phi_\beta}{R^2-\sum_\gamma\phi_\gamma\phi_\gamma}
\left[\frac{\sum_\gamma\phi_\gamma\partial^\mu\phi_\gamma+\sum_\delta\phi_\delta\partial_\mu\phi_\delta}
{R^2-\sum_\gamma\phi_\gamma\phi_\gamma}\right. \nonumber
\\ && \left.-\sum_\gamma\left(\partial^\mu\phi_\gamma\partial_\mu\phi_\gamma
+\phi_\gamma\Box \phi_\gamma\right)\right]=0 \label{llfc} \  .
\end{eqnarray}
$\alpha,\beta,\gamma=1,2$, $m_1^2=1 $, $m_2^2=\sigma^2 $. The ground
states are the homogeneous solutions of this system:
$\phi_1^0=\phi_2^0=0$, $\phi_3^0=\pm R$. In order to visualize these
configurations in, e. g., Figure 7 we draw the spin chain in such a
way that the $\phi_2:\phi_3$ plane is perpendicular to the $x$
spatial line whereas $\phi_1$ is aligned with the  $x$-axis. We
stress that this choice of basis is arbitrary but it is easy to
figure out the formulas and the graphics in another rotated basis
for the magnetization vector: $\vec{\phi}(x)=\phi_1(x)\vec{e}_1+
\phi_2(x)\vec{e}_2+\phi_3(x)\vec{e}_3=\phi_1^\prime(x)\vec{e^\prime}_1+
\phi_2^\prime(x)\vec{e^\prime}_2+\phi_3^\prime(x)\vec{e^\prime}_3$.
The main features of our preferred basis $\vec{e}_1, \vec{e}_2,
\vec{e}_3$ are: 1) The $\vec{e}_1$ vector points in the direction of
weaker $-V(\theta,\varphi)$ potential, see Figure 1(b). 2)
$\vec{e}_1, \vec{e}_2, \vec{e}_3$ is the basis used in the
continuous XY (in fact YZ) model of easy-axis ferromagnets near the
Curie point, see \cite{BW,BWZ} and References quoted therein.

\begin{figure}[htbp]
\centerline{\includegraphics[height=2.5cm]{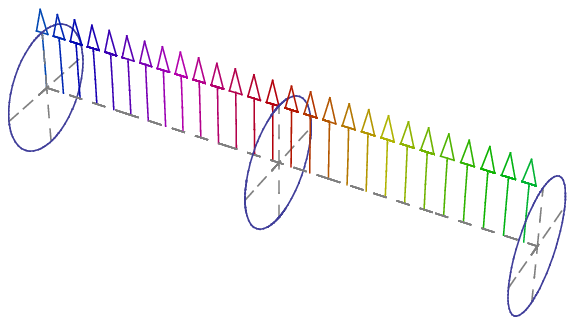}\quad \
\includegraphics[height=2.5cm]{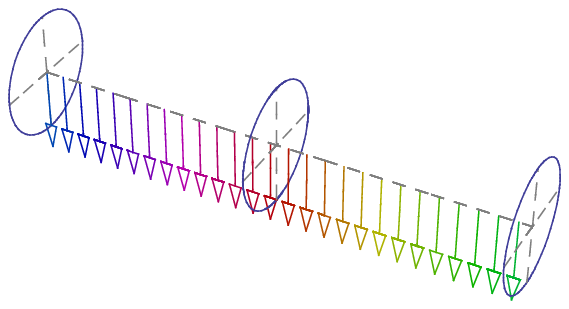}
} \caption{a) Ground state $\phi_3^0=R$ . All the spins are aligned
pointing to the North Pole b) Ground state $\phi_3^0=-R$. All the
spins are aligned pointing to the South Pole. }
\end{figure}

The spin fluctuations $\phi_1(t,x)=\delta\phi_1(t,x)$,
$\phi_2(t,x)=\delta\phi_2(t,x)$ around the ground state
$\phi_3(t,x)=R$ satisfy the linearized equations:
\begin{eqnarray*}
0&=&\frac{\partial\delta\phi_2}{\partial
t}+\frac{\partial^2\delta\phi_1}{\partial
t^2}-\frac{\partial^2\delta\phi_1}{\partial x^2}+\delta\phi_1
\\0&=& -\frac{\partial\delta\phi_1}{\partial
t}+\frac{\partial^2\delta\phi_2}{\partial
t^2}-\frac{\partial^2\delta\phi_2}{\partial x^2}+\sigma^2
\delta\phi_2^2\, .
\end{eqnarray*}
Therefore, the spin waves:
\begin{eqnarray}
\delta\phi_\alpha(t,x)&=&{1\over\sqrt{\lambda L}}\sum_k\,
{1\over\sqrt{\omega(k)}}\left(a_\alpha(k) e^{i\omega t- i k
x}+\right.\nonumber \\ && \left. +a_\alpha^*(k) e^{-i\omega t+ i k
x}\right)  \label{ssw}
\end{eqnarray}
satisfying periodic boundary conditions
$\delta\phi_\alpha(t,x)=\delta\phi_\alpha(t,x+\lambda L)$ are
solutions of (\ref{ssw}) for the frequencies complying with the
homogeneous system of algebraic equations:
\begin{equation}
\left(\begin{array}{cc} -\omega^2+k^2+1 & i\omega
\\ -i\omega & -\omega^2+k^2+\sigma^2 \end{array}\right)\left(\begin{array}{c} a_1(k)
\\ a_2(k)
\end{array}\right)=\left(\begin{array}{c} 0 \\ 0
\end{array}\right) \label{assw} \, .
\end{equation}
At the long wavelength limit $\omega^2<<\omega$, (\ref{assw}) is
tantamount to the non-relativistic dispersion law
\[
\omega^2(k^2)=(k^2+1)(k^2+\sigma^2)
\]
characteristic of ferromagnetic materials, although the quadratic
terms in the free energy prevent the standard $\omega(k)=k^2$ form.
\subsection{Bloch and Ising walls}

One may check that the $K_1/K_1^*$ kinks (\ref{k1kinks}) solve the
static Landau-Lifshitz equations (\ref{llfc}) on the $\phi_1=0$
orbit:
\[
\frac{d^2\phi_2}{dx^2}=\frac{-\phi_2}{R^2-\phi_2^2}\left[\frac{(\phi_2\frac{d\phi_2}{dx})^2}{R^2-\phi_2^2}
+\left(\frac{d\phi_2}{dx}\right)^2+\phi_2\frac{d^2\phi_2}{dx^2}\right]+\sigma^2\phi_2
\]
The $K_1/K_1^*$ kinks of the non-linear sigma model are consequently
solitary spin waves of this non-relativistic system, see Figure 8.
\begin{figure}[htbp]
\centerline{\includegraphics[height=2.8cm]{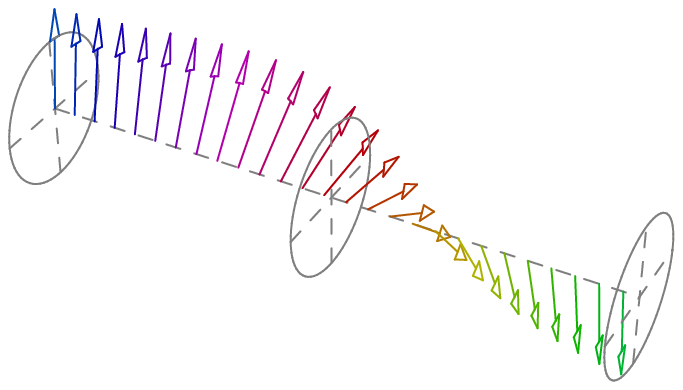}\
\includegraphics[height=2.8cm]{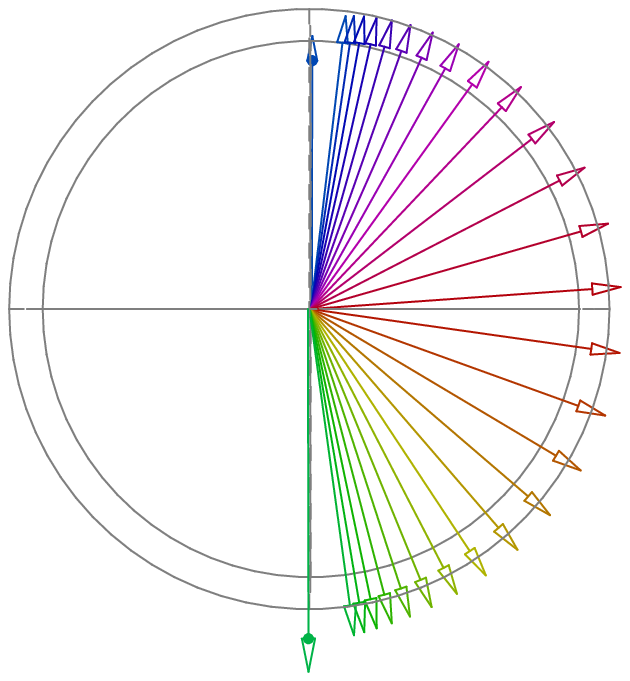}
} \caption {Graphic arrow representation of the $K_1$ kinks: {a)
$K_1$ spin chain. b) Perspective from one component of the boundary
of ${\mathbb S}^2\times{\mathbb R}$ showing how the spin flip
happens by means of a $\pi$-rotation around the $\phi_1$-axis.}}
\end{figure}

Simili modo, the $K_2/K_2^*$ kinks (\ref{k2kinks}) solve
(\ref{llfc}) along the $\phi_2=0$ kink orbit:
\[
\frac{d^2\phi_1}{dx^2}=\frac{-\phi_1}{R^2-\phi_1^2}\left[\frac{(\phi_1\frac{d\phi_1}{dx})^2}{R^2-\phi_1^2}
+\left(\frac{d\phi_1}{dx}\right)^2+\phi_1\frac{d^2\phi_1}{dx^2}\right]+\phi_1
\]
and are also spin solitary waves in this system, (Fig. 9).
\begin{figure}[htbp]
\centerline{\includegraphics[height=2.8cm]{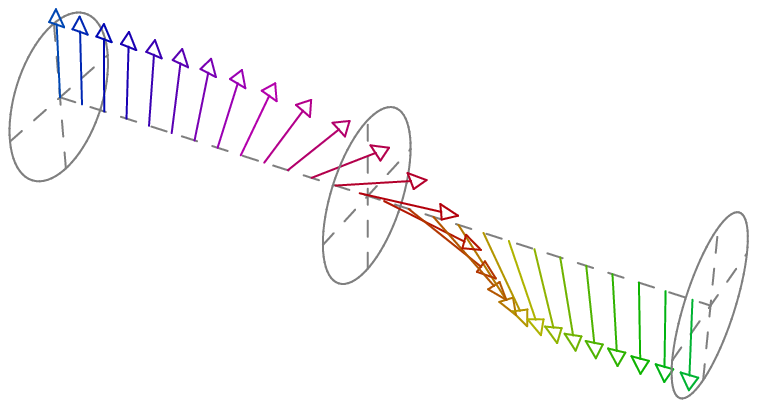}\
\includegraphics[height=2.8cm]{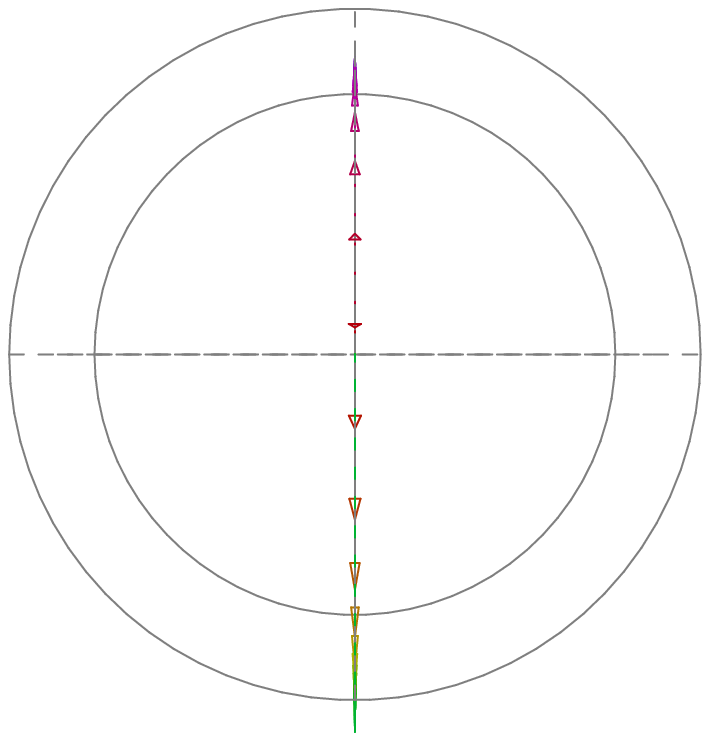}} \caption {Graphic arrow representation of the $K_2$ kink
{a) $K_2$ spin chain. b) Perspective from one component of the
boundary of ${\mathbb S}^2\times{\mathbb R}$ showing a forward spin
flip.}}
\end{figure}

Because the system of ODE's giving static solutions of the
({\ref{llf}) PDE system is the same as the static field equations of
the non-linear ${\mathbb S}^2$-sigma model, the NTK kinks are also
solitary spin waves, see Figure 10.
\begin{figure}[htbp]
\centerline{\includegraphics[height=2.8cm]{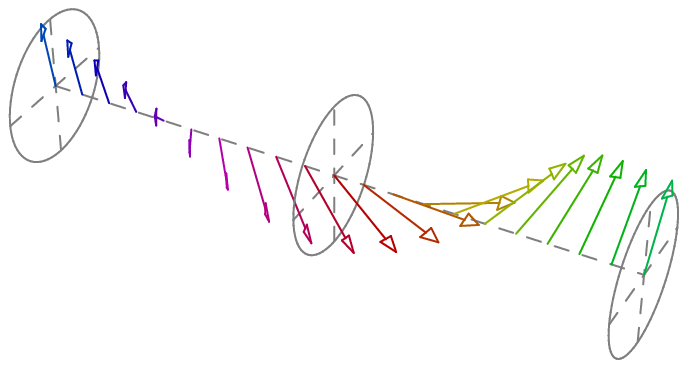}\
\includegraphics[height=2.8cm]{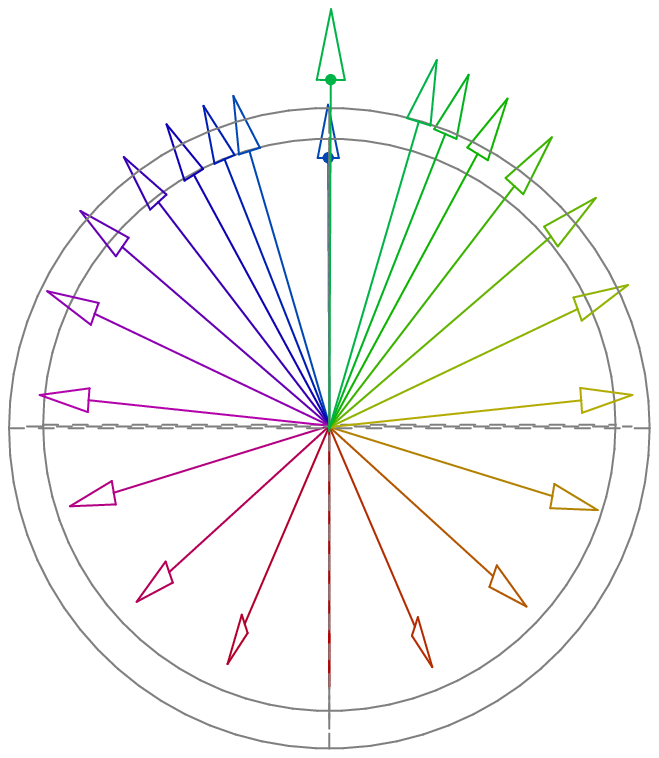}}
\caption  {Graphic arrow representation of $K_{\gamma_2}$ kinks: {a)
$K_{\gamma_2}$ spin chain. b) Perspective from the boundary of
${\mathbb S}^2\times{\mathbb R}$ showing the $2\pi$ rotation around
the $\phi_1$-axis of the spin to come back to the initial ground
state.}}
\end{figure}

In sum, understood as solitary spin waves $K_1/K_1^*$ kinks are
Bloch walls whereas $K_2/K_2^*$ kinks are Ising walls describing
interfaces between ferromagnetic domains, see \cite{BW}, \cite{BWZ}.
In this model we have thus found a moduli space of solitary waves
with an structure very similar to the structure of the space of
solitary waves of the XY model described in References \cite{BW} and
\cite{BW1}. There are Bloch and Ising walls and a one-parametric
family of NTK kinks that are non-linear superpositions of one Bloch
and one Ising wall with arbitrary separation between their centers.
The novelties here are: a) there is no need in the free energy of
fourth-order terms in the magnetization in the non-linear sigma
model for finding these mixtures of Bloch and Ising walls. b) The
analytical expressions (\ref{eq:ntksol}) differ from their analogues
in the XY model.

From the stability analysis performed in previous Sections, it is
clear that only the Bloch walls are stable and saturate the
Bogomolny bound. Things are different at the $\sigma=1$ limit where
all the kinks are topological, Bloch walls, and saturate the
Bogomolny bound. In this latter case the structure of the kink space
is akin to the kink space structure of the BNRT model \cite{BNRT},
see \cite{SV}, \cite{AMAJ1}, \cite{SAH}. There is a one-parametric
family of degenerate Bloch walls saturating the Bogomolny bound.

\section{Further comments: supersymmetry and stability}
Finally, we briefly explore the possibility of embedding our bosonic
model with its moduli space of kinks in a broader supersymmetric
framework. It turns out that the simpler ${\cal N}=1$, ${\it d}=1+1$
SUSY version of the massive non-linear ${\mathbb S}^2$-sigma model
only exists if the masses of the pseudo Nambu-Goldstone bosons are
equal ($\sigma=1$). It also seems difficult to build more exotic
possibilities coming from dimensional reduction of models of Kahler
or hyper-Kahler nature because the potential energy density is not
compatible with complex structures when $\sigma\neq 1$.

\subsection{Isothermal coordinates}
It is convenient to introduce isothermal coordinates in the chart
${\mathbb S}^2-\{(0,0,-R)\}$, which are obtained via stereographic
projection from the South Pole:
\begin{eqnarray}
\chi^1&=&\frac{\phi_1}{1+\frac{\phi_3}{R}}=\frac{R \phi_1}{R+{\rm
sg}(\phi_3)\sqrt{R^2-\phi_1^2-\phi_2^2}} \nonumber\\
\chi^2&=&\frac{\phi_2}{1+\frac{\phi_3}{R}}=\frac{R \phi_2}{R+{\rm
sg}(\phi_3)\sqrt{R^2-\phi_1^2-\phi_2^2}} \label{iso}\, .
\end{eqnarray}
The metric and the action in this coordinate system read:
\begin{eqnarray*}
&&ds^2=\frac{4
R^4}{(R^2+\chi^1\chi^1+\chi^2\chi^2)^2}(d\chi^1d\chi^1+d\chi^2d\chi^2)\\
&&S[\chi^1,\chi^2]=\int \, dx^2 \, \frac{2
R^4}{(R^2+\chi^1\chi^1+\chi^2\chi^2)^2}
\cdot\left[\partial_\mu\chi^1\partial^\mu\chi^1+\right. \\ &&\left.
+\partial_\mu\chi^2\partial^\mu\chi^2
-(\chi^1\chi^1+\sigma^2\chi^2\chi^2)\right] \ ,
\end{eqnarray*}
whereas the $K_1$ kinks are given by:
\begin{equation}
\chi^1_{K_1}(x)=0 \, , \    \chi^2_{K_1}(x)=\pm R \, \, {\rm
exp}[\pm\sigma (x-x_0)] \label{t1kplb} \ ,
\end{equation}
and we rewrite the second order fluctuation operator around the
$K_1$ kink (with $\chi_{K_1}^2(x)=Re^{-\sigma x})$) in the form:
\begin{eqnarray*}
&&\Delta_{K_1}\eta = -\left(\frac{d^2\eta^1}{dx^2}+2\sigma (1-{\rm
tanh}\sigma x)\frac{d\eta^1}{dx}
\right.\\&&\left.-\left(1-2\sigma^2+2\sigma^2\tanh\sigma
x\right)\eta^1\right)\frac{\partial}{\partial\chi^1}- \left( \frac{d^2\eta^2}{dx^2}+\right.\\
&& \left. +2\sigma (1-\tanh\sigma
x)\frac{d\eta^2}{dx}+\sigma^2\left(1-2\tanh\sigma
x\right)\eta^2\right)\frac{\partial}{\partial\chi^2} \ .
\end{eqnarray*}
In a parallel frame
$\mu=\mu^1(x)\frac{\partial}{\partial\chi^1}+\mu^2(x)\frac{\partial}{\partial\chi^2}\in\Gamma(T{\mathbb
S}^2\left|_{K_1})\right.$,
$\frac{d\mu^i}{dx}+\Gamma^i_{jk}(\chi_K)\chi_K^{\prime j}\mu^k=0$,
along the $K_1$ kink:
\begin{eqnarray*}
&&\frac{d\mu^1}{dx}+\sigma(1-\tanh)\mu^1(x)=0 \, \Rightarrow
\, \mu^1(x)=1+e^{-2\sigma x} \\
&&\frac{d\mu^2}{dx}+\sigma(1-\tanh)\mu^2(x)=0 \, \Rightarrow \,
\mu^2(x)=1+e^{-2\sigma x} \  .
\end{eqnarray*}
we recover the P$\ddot{\rm o}$sch-Teller operators:
\begin{eqnarray}
&&\Delta_{K_1}\eta =\left(
-\frac{d^2\eta^1}{dx^2}+(1-\frac{2\sigma^2}{{\rm cosh}^2\sigma
x})\eta^1\right)\, (1+e^{-2\sigma x})\frac{\partial}{\partial\chi^1}
\nonumber \\&& +\left(
-\frac{d^2\eta^2}{dx^2}+(\sigma^2-\frac{2\sigma^2}{{\rm
cosh}^2\sigma x})\eta^2\right)\, (1+e^{-2\sigma
x})\frac{\partial}{\partial\chi^2} \label{sodok1p}
\end{eqnarray}
Note that now the $K_1$ orbits are the positive and negative
ordinate half-axes, the stereographic projections of the
$\varphi=\frac{\pi}{2}$ and $\varphi=\frac{3\pi}{2}$ half-meridians,
such that fluctuations orthogonal to the orbit run in the direction
of the abscissa axis.

\subsection{The ${\cal N}=1$ massive SUSY sigma model}
In Reference \cite{AGMT} we analyzed the relationship of the
complete solution of the Hamilton-Jacobi equation for zero energy
and the superpotential of a supersymetric associated classical
mechanical system. Thus, we are tempted to use the Hamilton
characteristic function
\begin{eqnarray}
&&W^{(\beta_1,\beta_2)}(\chi)=\frac{(-1)^{\beta_1}R^2}{R^2+\chi^1\chi^1+\chi^2\chi^2}\cdot
\label{nlsup}\\ &&
\sqrt{\left(\sigma_+(\beta_2)R^2+\sigma_-(\beta_2)(\chi^1\chi^1+\chi^2\chi^2)\right)^2
-4\bar{\sigma}^2R^2\chi^1\chi^1} ,\nonumber
\end{eqnarray}
$\sigma_\pm(\beta_2)=1\pm (-1)^{\beta_2}\sigma$, to build the ${\cal
N}=1$ SUSY extension of our massive non-linear ${\mathbb S}^2$-sigma
model. On one hand we have that:
\[
\frac{1}{2}g^{ij}\frac{\partial
W^{(\beta_1,\beta_2)}}{\partial\chi^i}\cdot\frac{\partial
W^{(\beta_1,\beta_2)}}{\partial\chi^j}=\frac{2R^2(\chi^1\chi^1+\sigma^2\chi^2\chi^2)}{(R^2+\chi^1\chi^1+\chi^2\chi^2)^2}
\  ,
\]
$\forall \beta_1,\beta_2$. On the other hand (\ref{nlsup}) is free
of branch points only for $\sigma=1$. Supersymmetry does not allow
superpotentials with branch points and it seems that Hamilton-Jacobi
characteristic functions are compatible with a weaker form called
pseudo-supersymmetry in \cite{PT}. We close our eyes to this fact
for a moment and proceed to formally build the ${\cal N}=1$ SUSY
extension of our model using (\ref{nlsup}).

There are also two Majorana spinor fields:
\[
\psi^i(x^\mu)=\left(\begin{array}{c}\psi_1^i(x^\mu)\\
\psi_2^i(x^\mu)\end{array}\right)\, , \,
(\psi_\alpha^i)^*=\psi_\alpha^i ,\ \alpha=1,2\,.
\]
We choose the Majorana representation $\gamma^0=\sigma^2,
\gamma^1=i\sigma^1, \gamma^5=\sigma^3$ of the Clifford algebra
$\{\gamma^\mu,\gamma^\nu\}=2g^{\mu\nu}$ and define the Majorana
adjoints as: $\bar{\psi}^i=(\psi^i)^t \gamma^0$. The action of the
supersymmetric model is:
\begin{eqnarray*}
S=\int
\frac{dx^2}{2}\left\{g_{ij}\left(\partial_\mu\chi^i\partial^\mu\chi^j+
i\bar{\psi}^i\gamma^\mu(\partial_\mu\psi^j+\Gamma^j_{lk}\partial_\mu\chi^k\psi^l)\right)\right.&&
\\ \left.
-\frac{1}{6}R_{ijlk}\bar{\psi}^i\psi^j\bar{\psi}^l\psi^k-g^{ij}\frac{\partial
W}{\partial\chi^i}\frac{\partial
W}{\partial\chi^j}-\bar{\psi}^i\frac{D\partial
W}{\partial\chi^i\partial\chi^j}\psi^j \right\} \, \, ,&&
\end{eqnarray*}
where $\frac{D\partial
W}{\partial\chi^i\partial\chi^j}=\frac{\partial^2}{\partial\chi^i\partial\chi^j}
+\Gamma^k_{ij}\frac{\partial W}{\partial\chi^k}$. The spinor
supercharge
\begin{equation}
Q=\int \, dx \,
g_{ij}\left(\gamma^\mu\gamma^0\psi^i\partial_\mu\chi^j+i\gamma^0\psi^ig^{jk}\frac{\partial
W}{\partial\chi^k}\right)
\end{equation}
acts on the configuration space and leaves the action invariant.
Time-independent finite energy configurations complying with
\begin{equation}
\frac{d\chi^i}{dx}=g^{ij}\frac{\partial W}{\partial\chi^j} \ ,\
\psi^i_1(x)=-\psi^i_2(x) \label{susk}
\end{equation}
annihilates the supercharge combination $Q_1+Q_2$ and these
solutions might be interpreted as $\frac{1}{2}$ BPS states in this
supersymmetric framework. In particular, the SUSY $K_1$ kinks
\begin{eqnarray*}
\chi^1_{K_1}(x)=0  &,&  \chi^2_{K_1}=\pm
R e^{\pm\sigma x} \\
\psi^1_{K_1}(x)=\left(\begin{array}{c} 0 \\ 0 \end{array}\right)
 &,&  \psi^2_{K_1}(x)=\pm\sigma R e^{\pm\sigma x}\left(\begin{array}{c} 1 \\
-1
\end{array}\right)
\end{eqnarray*}
satisfy (\ref{susk}) (with appropriate choices of $\beta_1$,
$\beta_2$). Note that $\psi^2_{K_1}(x)$ is the SUSY partner of
$\chi^2_{K_1}(x)$ under the action of the broken SUSY supercharge
$Q_1-Q_2$. We also remark that
\[
\frac{d\chi^2_{K_1}}{dx}=\pm\sigma R e^{\pm\sigma x}=\pm\sigma R
(1+e^{\pm 2\sigma x})\cdot\frac{1}{\cosh\sigma x} \   ,
\]
i.e., the fermionic partner in the SUSY kink is the zero mode of the
second order fluctuation operator back from the parallel frame to
the $K_1$ orbit.

\subsection{Fermionic fluctuations}

The Dirac equation ruling the small fermionic fluctuations on the
$K_1$ kink reads:
\begin{eqnarray}
D
\delta\psi^i(t,x)&=&i(\gamma^0\partial_0-\gamma^1\partial_1)\delta\psi^i(t,x)\nonumber
\\ &&
-i\gamma^1\Gamma^i_{jk}(\chi_{K_1})\partial_1\chi^j_{K_1}(x)\delta\psi^k(t,x)\nonumber\\
&&+ g^{ij}(\chi_{K_1})\frac{D\partial
W}{\partial\chi^j\partial\chi^k}(\chi_{K_1})\delta\psi^k(t,x)\quad .
\label{dir}
\end{eqnarray}
Acting on (\ref{dir}) with the adjoint Dirac operator, the search
for solutions of $D^\dagger D \delta\psi^i(t,x)=0$ of the stationary
form $\delta\psi^i(t,x)=e^{i\omega t}\delta\varrho^i(x,\omega)$
requires us to deal with the following ODE system:
\begin{eqnarray*}
&&-\frac{d^2}{dx^2}\delta\varrho^i(x)+g^{ij}\frac{D\partial
W}{\partial\chi^j\partial\chi^k}\cdot g^{kl}\frac{D\partial
W}{\partial\chi^l\partial\chi^m}\delta\varrho^m(x)\\ &&
+R^i_{jkl}\frac{d}{dx}\chi^j_{K_1}
\frac{d}{dx}\chi^k_{K_1}\delta\varrho^l(x)\\ &&-i \gamma^1
g^{ij}\frac{\partial W}{\partial\chi^j}\cdot g^{kl}\frac{D^2\partial
W}{\partial\chi^k\partial\chi^l\partial\chi^m}\delta\varrho^m(x)=\omega^2\delta\varrho^i(x)
\end{eqnarray*}
valued at $\chi=\chi_{K_1}$.

On eigenspinors of $-i\gamma^1=\sigma^1$,
$\delta\varrho^i_1(x)=\pm\delta\varrho^i_2(x)=\delta\varrho^i_\pm(x)$,
the above spectral ODE system reduce to the (symbolically written)
pair of equations:
\begin{equation}
\bigtriangleup^{\pm}_{K_1}\delta\varrho_\pm=\left[-\frac{d^2}{dx^2}+W^{\prime\prime}\otimes
W^{\prime\prime}+R\pm W^{\prime}\otimes
W^{\prime\prime\prime}\right]\delta\varrho_\pm \label{ffsp} \   .
\end{equation}
$\bigtriangleup^+_{K_1}$ is exactly equal to the second order
differential operator ruling the bosonic fluctuations. Therefore, in
the parallel frame to the $K_1$ orbit we write
$\bigtriangleup^+_{K_1}$ in matrix form:
\[
\bigtriangleup^+_{K_1}=\left(\begin{array}{cc}-\frac{d^2}{dx^2}+1-\frac{2\sigma^2}{{\rm
cosh}^2\sigma x}& 0 \\ 0 &
-\frac{d^2}{dx^2}+\sigma^2-\frac{2\sigma^2}{{\rm cosh}^2\sigma
x}\end{array}\right) \,.
\]
In the same frame $\bigtriangleup^-_{K_1}$ is the intertwined
partner, see \cite{MRNW}:
\[
\bigtriangleup^-_{K_1}=\left(\begin{array}{cc}-\frac{d^2}{dx^2}+1& 0
\\ 0 & -\frac{d^2}{dx^2}+\sigma^2\end{array}\right) \, .
\]
If $\sigma\neq 1$, there is a bound state in
$\bigtriangleup^+_{K_1}$ of energy $1-\sigma^2$ unpaired with an
eigenstate of the same energy in $\bigtriangleup^-_{K_1}$, a fact
incompatible with supersymmetry as we expected from the use of the
complete solution of the Hamilton-Jacobi equation as superpotential,
closing our eyes to the fact that, related to the instability of
$NTK$ and $K_2$ kinks,  the Hamilton characteristic function has
branching points at the foci defining the elliptic coordinate
system. A similar problem arouse in \cite{ABGLM1} and \cite{ABGLM2}
where meromorphic Hamilton characteristic functions have been found.
It is an open problem to explore whether or not these milder
singularities allow the use of these Hamilton characteristic
functions as superpotentials to extend the bosonic models dealt with
in \cite{ABGLM1}, \cite{ABGLM2} to the supersymmetric framework.

If the masses are equal ($\sigma=1$), however, the Hamilton
characteristic function is free of branching points and the unpaired
states are zero modes. The ${\cal N}=1$ SUSY model is correct and we
can apply the SUSY version of the Cahill-Comtet-Glauber formula
proposed in \cite{BC1} to find the same one-loop correction to the
SUSY ${\mathbb S}^2$ kink as given in \cite{MRNW}:
\[
\bigtriangleup E_{K_1}^{\rm
SUSY}(\sigma=1)=-\frac{\lambda}{2\pi}\sum_{i=1}^2(\sin\nu_i^+-\nu_i^+\cos\nu_i^+)=-\frac{\lambda}{\pi}
\, \, .
\]
Here $\nu_1^+=\nu_2^+=\arccos (0)=\frac{\pi}{2}$ are the angles
obtained from the bound states of $\bigtriangleup_{K_1}^+$. There
are no bound states in the spectrum of $\bigtriangleup_{K_1}^-$.

\section{Acknowledgements}

We are grateful to M. Santander, S. Woodford, I. Barashenkov, M.
Nitta, P. Letelier, D. Bazeia, and Y. Fedorov for informative and
illuminating electronic/ordinary mail correspondence and/or oral
conversations on several issues concerning this work. Any
misunderstanding is the authors's own responsibility.

We also thank the Spanish Ministerio de Educacion y Ciencia and
Junta de Castilla y Leon for partial support under grants
FIS2006-09417 and GR224.


\begin{thebibliography}{99}

\bibitem{AMAJ} A. Alonso Izquierdo, M. A. Gonzalez Leon, and J. Mateos Guilarte,
Phys. Rev. Lett. {\bf 101} (2008) 131602

\bibitem{CCG} K. Cahill, A. Comtet, and R. Glauber, Phys. Lett. {\bf
64B}(1976) 283-285

\bibitem{AT} E. R. C. Abraham and P. K. Townsend, Phys. Lett. {\bf
B291} (1992) 85-88

\bibitem{AT1} E. R. C. Abraham and P. K. Townsend, Phys. Lett. {\bf
B295} (1992) 225-232

\bibitem{Nit} M. Arai, M. Naganuma, M. Nitta, and N. Sakai, Nucl.
Phys. {\bf B652} (2002) 35-71

\bibitem{Dor} N. Dorey, JHEP 9811 (1998) 005

\bibitem{Lee} R. A. Leese, Nucl. Phys. {\bf B366} (1991) 283-314

\bibitem{Abr} E. Abraham, Phys. Lett. {\bf B278} (1992) 291-296

\bibitem{GPTT} J. P. Gauntlet, R. Portugues, D. Tong, P. K.
Townsend, Phys. Rev. {\bf D63} (2001) 085002

\bibitem{INOS} Y. Isozumi, M. Nitta, K. Oshasi, N. Sakai, Phys. Rev.
{\bf D71} (2005) 065018

\bibitem{EINOS} M. Eto, Y. Isozumi, M. Nitta, K. Oshasi, N. Sakai,
Jour. Phys. {\bf A39} (2006) R315-R392

\bibitem{ES} M. Eto, and N. Sakai, Phys. Rev. {\bf D68} (2003)
125001

\bibitem{BG} D. Bazeia, and A. R. Gomes, JHEP {\bf 05}(2004) 012

\bibitem{SAH} A. de Souza Dutra, A. C. Amaro de Faria, and M. Holt,
Phys. Rev. {\bf D78}(2008) 043526

\bibitem{Col} S. F. Coleman, Comm. Math. Phys. {\bf 31} (1973) 259

\bibitem{Col1} S. F. Coleman, {\it \lq\lq Aspects of symmetry"}, Cambridge
University Press, 1985, Chapter 6: {\sl \lq\lq Classical lumps and
their quantum descendants"}

\bibitem{BC} L. J. Boya, and J. Casahorran, Ann. Phys. {\bf 196}
(1989) 361-385

\bibitem{MRNW} C. Mayrhofer, A. Rehban, P. van Nieuwenhuizen, and R.
Wimmer, JHEP (2007) 0709:069

\bibitem{AMAWJ1} A. Alonso Izquierdo, W. Garcia Fuertes, M. A. Gonzalez Leon, and J.
Mateos Guilarte, Nucl. Phys {\bf B 638} (2002) 378-404

\bibitem{AMAWJ2}  A. Alonso Izquierdo, W. Garcia Fuertes, M. A. Gonzalez Leon, and J.
Mateos Guilarte, Nucl. Phys {\bf B 635} (2002) 525-557

\bibitem{AMAWJ3}  A. Alonso Izquierdo, W. Garcia Fuertes, M. A. Gonzalez Leon, and J.
Mateos Guilarte, Nucl. Phys {\bf B 681} (2004) 163-194

\bibitem{AJ} A. Alonso Izquierdo, and J. Mateos Guilarte, Physica
{\bf D 237} (2008)3263-3291

\bibitem{CPNS} S. F. Coleman, S. Park, A. Neveu, and C. Sommerfield,
Phys. Rev. {\bf D15} (1977) 544

\bibitem{MF} P. Morse and H. Feshbach, {\it \lq\lq Methods of Theoretical
Physics"}, Volume I, McGraw Hill, New York, 1953

\bibitem{Dub} B. Dubrovine, Russ. Math. Surv. {\bf 36:2}(1981)11-80

\bibitem{Bog} E. Bogomolny, Sov. J. Nucl. Phys.{\bf 24} (1976) 449

\bibitem{Per}  A. Perelomov, {\it Integrable Systems of Classical
Mechanics and Lie Algebras}, Birkhauser, (1992)

\bibitem{Ito} H. Ito, Phys. Lett. {\bf 112A} (1985) 119

\bibitem{AAi} A. Alonso Izquierdo, M. A. Gonzalez Leon, J.
Mateos Guilarte, J. Phys. {\bf A31} (1998) 209

\bibitem{ItoT} H. Ito, H. Tasaki, Phys. Lett. {\bf A113} (1985)
179

\bibitem{MG} J. Mateos Guilarte, Ann. Phys. {\bf 188} (1988) 307

\bibitem{Hal} F. D. M. Haldane, Phys. Rev. Lett. {\bf 50} (1983) 1153

\bibitem{BW} S. R. Woodford, and I. V. Barashenkov, J. Phys. A: Math.
Teor. {\bf 41} (2008) 185203

\bibitem{BW1} S. R. Woodford, and I. V. Barashenkov, Phys. Rev. {\bf
E 75} (2007) 026605

\bibitem{BWZ} I. V. Barashenkov, S. R. Woodford, and E. V.
Zemlyanaya, Phys. Rev. Lett. {\bf 90} (2003) 054103

\bibitem{BNRT} D. Bazeia, J. R. Nascimento, R. Ribeiro, and D.
Toledo, J. Phys. A: Math. Gen. {\bf 30}(1997) 8157

\bibitem{SV} M. A. Shifman, and M. B. Voloshin, Phys. Rev. {\bf D57}
(1998) 2590

\bibitem{AMAJ1} A. Alonso Izquierdo, M. A. Gonzalez Leon, J.
Mateos Guilarte, Phys. Rev. {\bf D65} (2002) 085012

\bibitem{AMAJ2} A. Alonso Izquierdo, M. A. Gonzalez Leon, J.
Mateos Guilarte, Nonlinearity {\bf 15} (2002) 1097

\bibitem{AGMT} A. Alonso Izquierdo, M. A. Gonzalez Leon, J. Mateos
Guilarte, and M. de la Torre Mayado, Ann. Phys. {\bf 308} (2003)
664-691

\bibitem{PT} P. K. Townsend, Class. Quant. Grav. {\bf 25}
(2008)045017

\bibitem{ABGLM1} V. Afonso, D. Bazeia, M. A. Gonzalez Leon, L.
Losano, and J. Mateos Guilarte, Phys. Lett. {\bf B662}(2008)74-79

\bibitem{ABGLM2} V. Afonso, D. Bazeia, M. A. Gonzalez Leon, L.
Losano, and J. Mateos Guilarte, Nucl. Phys. {\bf B810}[FS]
(2009)427-459

\bibitem{BC1} L. J. Boya, and J. Casahorran, Jour. Phys. {\bf A23}
(1990) 1645

\end{thebibliography}
\end{document}